\documentclass[11pt,a4paper]{article}

\parskip \bigskipamount

\usepackage{graphicx}
\usepackage{amssymb}
\usepackage{amsmath}
\usepackage{bm}
\usepackage{epsf}
\usepackage{mdframed}
\usepackage{authblk}

\begin{document}

\title{Quantum Electronics for Fundamental Physics}

\author[1]{Stafford Withington}

\affil[1]{Department Physics, University of Oxford, Oxford OX1 3PU, UK}

\maketitle

\begin{abstract}
The emerging field of quantum sensors and electronics for fundamental physics is introduced, emphasising the role of thin-film superconducting devices. Although the next generation of ground-based and space-based experiments requires the development of advanced technology across the whole of the electromagnetic spectrum, this article focuses on ultra-low-noise techniques for radio to far-infrared wavelengths, where existing devices fall short of theoretical limits. Passive circuits, detectors and amplifiers are described from classical and quantum perspectives, and the sensitivities of detector-based and amplifier-based instruments discussed. Advances will be achieved through refinements in existing technology, but innovation is essential. The needed developments go beyond engineering and relate to theoretical studies that bring together concepts from quantum information theory, quantum field theory, classical circuit theory, and device physics. This article has been written to introduce graduate-level scientists to quantum sensor physics, rather than as a formal review.
\end{abstract}

\noindent {\bf In memory of my sister Diana Syder, clinical speech and language specialist, artist and poet, who dedicated her life to helping others communicate.}

\section{Quantum sensing}

Early in the 20th Century a series of experiments revealed that all waves exhibit particle-like behaviour, all particles exhibit wave-like behaviour, and that these phenomena are associated with an intrinsic indeterminacy in the outcomes of measurements. The nature of physical reality was questioned, and the language of classical mechanics was replaced by the formalism of quantum mechanics. It is now accepted that no matter how great the skill of an observer, the outcome of a single measurement on any simple physical system of any basic physical quantity is profoundly uncertain. When describing how systems evolve with time, we are forced into a new mechanics where probability distributions evolve in a deterministic manner, rather than the dynamical variables themselves. When several variables are measured, either simultaneously or sequentially, conditional probabilities come into play, and experimental measurements of one kind can influence those of another without any classical interactions being present. No physical quantity can be regarded as having an {\em actual} value until a measurement is made. This way of thinking is not merely a rebranding of classical statistical physics, where it is not humanly possible to keep track of every microscopic degree of freedom, such as the motion of every water molecule in a steam engine, but is intrinsic to the way in which we gather information about the physical world. Should scientists have ever gone down the quantum rabbit hole?  Well, quantum mechanics is not optional, but is essential if we are to build mathematical models that replicate the behaviour of experimental systems.

Quantum mechanics applies to all dynamical variables (not merely the mechanics of elementary particles), and therefore it applies to electrical quantities such as voltage, current, power, electric and magnetic fields, and dipole moments, etc. When devices and circuits are operated at low physical temperatures (10~mK to 4~K) to minimise thermal noise, the mysterious world of quantum mechanics is revealed, and it becomes {\em necessary} to use the techniques of quantum mechanics to describe the behaviour of electrical circuits. Because of the need to track probability distributions, the analysis of circuit elements, such as transformers, transmission lines, power detectors, mixers and amplifiers becomes complicated, and one is forced into asking questions about the influence of quantisation (vacuum fluctuations, squeezing, back action and entanglement) on the fidelity and sensitivity of electrical measurements.

From a measurement perspective there is some target that we wish to probe, and this target must be described quantum mechanically. Likewise, there is a sensor, which is usually part of a larger electrical circuit, which itself must be described quantum mechanically. The purpose of the sensor is to create a macroscopic quantity that can be recorded, and which carries faithful information about properties of the target. {\em Quantum Sensing} refers to the quantisation of the dynamical variables of the target and the interaction with the quantum behavior of the instrument carrying out the measurement. This interaction is potentially complicated because of the need to minimise the variance of the recorded signal, taking into account quantum uncertainty, and the inevitable quantum disturbance caused by the sensor. The quantum states of the target and sensor evolve in time in a mutually interactive way, which is the essence of quantum sensing.

This article introduces the emerging field of quantum sensors and electronics for fundamental physics. The work is first put in context by describing a number of fundamental problems in physics where ultra-low-noise experiments are needed. Measurement principles are then discussed, focusing on electromagnetic fields in the microwave (30 cm, 1 GHz) to far-infrared (30 $\mu$m, 10 THz) wavelength range, where the transition from wave-like to particle-like behaviour is most pronounced. Special consideration is given to passive circuits, power detectors and signal amplifiers, but the principles are closely related to other kinds of quantum measurement such as stress, strain, speed and orientation. Towards the end of the article, the relative sensitivities of power detectors and amplifiers are compared, and a number of ultra-low-noise superconducting devices are described. Throughout the article two points are emphasised: (i) current experiments fall short of theoretical sensitivity limits, and a new generation of technology is needed to push down into the quantum-dominated regime; (ii) considerable innovation is possible by bringing together concepts from quantum information theory, quantum field theory, classical circuit theory, and device physics into a mathematical framework that can be used for modelling.

\section{Why are ultra-sensitive measurements needed?}

A new generation of fundamental physics experiments requires access to a family of ultra-low-noise sensors that can operate over the radio, microwave and far-infrared regions of the electromagnetic spectrum. For example, there is a need to produce an all-sky map the polarisation state of the Cosmic Microwave Background Radiation (CMBR) to one part in 10$^{9}$ to understand the role of gravitational waves in forming structure in the early Universe \cite{Aba1,Xu1}. There is a need to observe the most distance galaxies (z$>$5) to understand how galaxies first formed and evolved, and arrived at the local universe we see today. There is a need to understand the nature of Dark Matter (DM), with one strongly motivated possibility being the existence of a family of low-mass particles ($\mu$eV to meV)  that interact only weakly with electromagnetic fields \cite{Ant1,Sik1}. Determining the absolute mass of the neutrino remains one of the most pressing problems in laboratory physics, and an international effort is underway to understand how it can be achieved by measuring, to one part in 10$^{6}$,  the energies of individual electrons released during the radioactive decay of Tritium \cite{Ash1,Asn1}. There is a need for laboratory experiments that can probe the nature of spacetime and its relationship with the fundamental postulates of quantum field theory. These experiments, and others, require sensors that push at the limits imposed by quantum mechanics.

\section{Power detection - a classical perspective}

\label{sec_pow_det}

At long wavelengths, scientists measure power, both dissipative and reactive, whereas at short wavelengths, they measure photon rates, and if possible count photons. Whilst it is true that average power $P$ is related to average photon rate $W$ by $P = \hbar \omega W$, the quantisation of the electromagnetic field goes well beyond this simple equality, and on moving from microwave to optical wavelengths, the behaviour of low-noise instruments changes significantly.

\subsection{Radiation and noise}

\label{sec_rad_nse}

At long wavelengths, waveguiding systems tend to be {\em single mode} meaning that there is a single spatial transverse degree of freedom available for carrying power from the source to the detector. At infrared and optical wavelengths, free-space beams tend to be highly {\em multimode}, meaning that there is a large number of transverse degrees of freedom available. By counting modes in wavevector space ($k$-space), the longitudinal mode rate $J$ can be calculated for (i) a transverse electromagnetic (TEM) transmission line, (ii) radiation into a half space, and (iii) radiation into solid angle $\Omega = 4 \pi \sin^{2} (\theta_{\rm m}/2)$ where $\theta_{\rm m}$ is the half opening angle of the beam: Table~\ref{modes}. Because most modes travel obliquely to the assumed optical axis, it is necessary to weight each mode by a projected area, giving an effective solid angle $\Omega_{\rm eff} = \pi \sin^{2} (\theta_{\rm m})$: $\Omega_{\rm eff} \rightarrow \Omega$ as $\theta_{\rm m} \rightarrow 0$. The overall differential modal flux can then be written ${\rm d}J = N {\rm d} \omega / 2 \pi$, where $N = A \Omega / \lambda^{2}$ is the effective number of transverse modes (not including polarisation). The effective number of transverse modes can be calculated rigorously using diffraction theory, and the above expression for $N$ is accurate even when the throughput is small and the transmission spectrum tapers off gradually with mode number. The modal structure of fields is classical, but it is important to appreciate that each mode constitutes a degree of freedom, which must be quantised accordingly.

\begin{table}
\begin{center}
{\begin{tabular}{cccc}
  & Mode Rate & Effective Mode Rate & Number of Modes \\ \hline
  TEM line & ${\rm d}J = \frac{1}{2 \pi} {\rm d} \omega$ & $ {\rm d} J = \frac{1}{2 \pi} {\rm } {\rm d} \omega$ & 1  \\ \hline
  Half Space & ${\rm d}J = \frac{A \omega^{2}}{4 \pi^{2} c^{2}} {\rm d} \omega $& ${\rm d} J = \frac{A \omega^{2}}{8 \pi^{2} c^{2}} {\rm d} \omega $ & $N = \frac{1}{2} \frac{A \Omega}{\lambda^{2}} =  \frac{A \Omega_{\rm eff}}{\lambda^{2}}$ \\ \hline
  Solid Angle $\Omega$ & $ {\rm d}J = \frac{A \omega^{2} \Omega}{8 \pi^{3} c^{2}} {\rm d} \omega  $ & $ {\rm d} J = \frac{A \omega^{2} \Omega_{\rm eff}}{8 \pi^{3} c^{2}} {\rm d} \omega  $ & $N = \frac{A \Omega_{\rm eff}}{\lambda^{2}}$ \\ \hline
\end{tabular}}
\end{center}
\label{modes}
\caption{Longitude mode rate ${\rm d}J$, effective longitudinal mode rate after projecting onto a plane, and effective number of transverse modes $N$ for a TEM transmission line, radiation into a half space and radiation into physical solid angle $\Omega$. $A$ is the area of the source, and $\Omega_{\rm eff}$ is the effective solid angle of the beam.
}
\end{table}

If each mode, comprising both longitudinal and transverse parts, is quantised and in a thermal state at temperature $T_{\rm p}$, the average power $P$ and its variance $(\Delta P)^{2}$ become
\begin{align}
\nonumber
P & =  \frac{A \Omega_{\rm eff}}{\lambda^{2}} \int \frac{{\rm d} \omega}{2 \pi} \, \frac{\hbar \omega}{e^{\hbar \omega / k T_{\rm p}} - 1} \\
\label{A1}
& = \int {\rm d} \nu \, P(\nu) \\ \nonumber
(\Delta P)^{2} & = \frac{1}{\tau} \frac{A \Omega_{\rm eff}}{\lambda^{2}} \int \frac{{\rm d} \omega}{2 \pi} \,   \frac{ ( \hbar \omega )^{2} e^{\hbar \omega / k T_{\rm p}}}{ \left( e^{\hbar \omega / k T_{\rm p}} -1 \right)^{2}} \\
\label{A2}
& = \frac{1}{\tau} \int {\rm d} \nu \, \left[ \frac{P(\nu)^{2}}{N} + h \nu P(\nu) \right],
\end{align}
where $P(\nu) = N h \nu n(\nu)$ is the average spectral power, $n(\nu)$ is the single-mode thermal occupancy, and $\tau$ is the time over which energy is collected to give the recorded power.  If a perfectly matched planar detector is illuminated by a thermal field, these expressions give the average power and fluctuations in power measured. Although $P$ can be subtracted from the recorded output to reveal any additional signal, the fluctuations are troublesome, and act as a noise source in addition to any noise generated by the detector itself. The first term in Equation (\ref{A2}) reduces to the radiometer equation $\Delta T_{\rm p} = T_{\rm p} / \sqrt{B_{\rm pre} \tau}$ for a single-mode detector when $\hbar \omega < k T_{\rm p}$, where $B_{\rm pre}$ is the pre-detection bandwidth. $\Delta T_{\rm p}$ is the amount by which the source temperature must change to produce an output that is discernable above the noise. Notice that $B_{\rm pre} \tau \approx \tau / \tau_{\rm c}$ is the number of independent samples having coherence time $\tau_{\rm c}$ in the integration period $\tau$. The first term in Equation (\ref{A2}) can be regarded as coming from Gaussian distributed fluctuations in the envelope of a classical wave. The second term is dominant when $\hbar \omega > k T_{\rm p}$, and has a variance that is proportion to the mean, which is characteristic of Poisson statistics. It is indicative of the photon counting statistics of a coherent quantum state, where the variance in occupancy is equal to the mean. Thus, classical fluctuations tend to dominate at low frequencies and quantum fluctuations at high frequencies, but in general they appear together and add as uncorrelated fluctuations, which is suggestive of different physical origins. For a 10 mK source, the changeover occurs at about 200 MHz.

This blended behaviour can also be understood in terms of a Poisson mixture comprising an ensemble of Poisson distributions having continuously distributed means. Assume that there is a single transverse mode, for example a TEM transmission line. Suppose that the individual longitudinal modes, each lasting about $1/{\rm d} \nu$, are in coherent states, but that the complex amplitudes $\alpha$ vary. By definition, a coherent state is an eigenstate of the annihilation operator, $\hat{a} | \alpha \rangle = \alpha | \alpha \rangle$; it corresponds mostly closely to a coherent classical wave having complex amplitude $\alpha$. $\hat{a}$ is not Hermitian, and so does not represent a directly-measurable single quantity, unlike the in and out of phase components. If $P(\xi)$ is the probability distribution of $\xi \equiv |\alpha^{2}|$,  where for a coherent state $|\alpha|^{2}$ is also the average occupancy $\langle n \rangle$. The probability of detecting $n$ photons in a longitudinal mode is given by the conditional probability $P(n|\xi)$, but because we are interested in the probability of detecting $n$ photons over the whole ensemble, equivalently over a long integration time $\tau > 1/{\rm d} \nu$,
\begin{equation}
\label{A3}
P(n) = \int {\rm d} \xi \, P(n|\xi) P(\xi).
\end{equation}
Using $E[n] = \sum n P(n)$ for the expectation value, and remembering that for a Poisson distribution $E[n|\xi] = \xi$, it can be shown using straightforward algebra that
\begin{equation}
\label{A4}
E[n] = E[\xi] = E[|\alpha|^{2}],
\end{equation}
which reproduces Equation (\ref{A1}). The average power in a wave having a randomly varying amplitude gives the average occupancy of the underlying Poisson process.

More interestingly, the variance $V[n] = E[n^{2}] - (E[n])^{2}$ becomes
\begin{align}
\label{A5}
\nonumber
V[n] & =  V[|\alpha|^{2}] + E[|\alpha|^{2}] \\
& = V [|\alpha|^{2}] + V[n|E[|\alpha|^{2}]].
\end{align}
The variance of the occupancy is the sum of the variance of $|\alpha|^{2}$, the classical noise, and the variance of a pure Poisson process  $V[n|E(|\alpha|^{2})]$, the quantum noise, having $E[|\alpha|^{2}]$ as its parameter. If the power is averaged for time $\tau$, the variance in the observations is
\begin{align}
\label{A6}
\nonumber
(\Delta P)^{2} & = \frac{1}{\tau} \int d \nu \, N (\hbar \omega)^{2} V(n) \\
& =  \frac{1}{\tau} \int d \nu \, N (\hbar \omega)^{2}  V [|\alpha|^{2}] \\ \nonumber
& + \frac{1}{\tau} \int d \nu \, N (\hbar \omega)^{2} V \left[ n|E[|\alpha|^{2}] \right].
\end{align}

When trying to progress, the first term of Equation (\ref{A6}) is awkward, because $V [|\alpha|^{2}]$ is needed, which depends on the unspecified distribution $P(\xi)$. Assume, in the spirit of the central limit theorem, that the quadrature components of the complex amplitude $\alpha$ are Gaussian variates. $P(\xi)$ is then a chi-squared distribution with one degree of freedom. A Gaussian distribution, however, has the feature that all of its moments can be calculated from the first and second moments. So, a more elegant approach is to say that $V [|\alpha|^{2}] = E[\alpha \alpha^{\ast} \alpha \alpha^{\ast}] - (E[\alpha \alpha^{\ast}])^{2}$, and then use the moment theorem (Isserlis' theorem) for complex Gaussian random processes \cite{Ree1,Iss1} to give $V [|\alpha|^{2}] =(E[|\alpha|^{2}])^{2} = n(\nu)^{2}$. The second term of Equation (\ref{A6}) is more straightforward because the variance is that of a pure Poisson distribution, $V \left[ n|E[|\alpha|^{2}] \right] = n(\nu)$, and so overall
\begin{align}
\label{A7}
(\Delta P)^{2} & =  \frac{1}{\tau} \int {\rm d} \nu \, \left[ \frac{P(\nu)^{2}}{N} + h \nu P(\nu) \right],
\end{align}
which reproduces the classical and quantum noise terms in Equation (\ref{A2}). An electromagnetic wave exhibits both wave-like and particle-like behaviour. Thus, heuristically, and with great caution, the image is that of a `blizzard' of particulate photons having an inhomogeneous density distribution. Photon counting statistics can be used to characterise different kinds of bunching at low light levels.

At long wavelengths, photon energies are small (4 $\mu$eV at 1 GHz) whereas at short wavelengths they are large (40 meV at 10 THz). The photon rate in a wave carrying fixed power increases significantly as the frequency is falls, making it difficult to distinguish individual events. Generally speaking, as the frequency increases, a beam accrues more transverse modes and the quantum statistics changes, leading to rich and complex behaviour. At long wavelengths, background-limited detectors are characterised in terms of average photon flux and temporal variations in the flux, whereas at short wavelengths, they are characterised in terms of photon counting statistics and average dark count rate.

\subsection{Multimode power detectors}

\label{sec_mul_det}

Section (\ref{sec_rad_nse}) assumes that the output of a detector replicates the statistical behaviour of the power at its input. When a detector can absorb energy through a number of transverse modes simultaneously, the situation is more complicated because it will in general respond differently to each of the modes. The characteristics of multimode classical fields are best described by second-order correlation functions, which for convenience can be written in terms of dyads $\overline{\overline{E}}({\bf r}_{1},t_{1};{\bf r}_{2},t_{2}) \equiv {\rm E} \left[ {\bf E} ({\bf r}_{1},t_{1}) {\bf E}({\bf r}_{2},t_{2}) \right]$, where  ${\bf E}({\bf r}_{1},t)$ is the vector valued electric field at space-time point $({\bf r},t)$. Correlation functions can be written using tensor or matrix notation, but dyadic algebra \cite{Mor1} is common in electromagnetism, and elasticity, and it is particularly convenient for vector-valued correlations, not least because of its similarity with scalar expressions. In paraxial systems, energy is assumed to flow with respect to some optical axis, $z$, and the field vectors are taken to be transverse, and so two dimensional.

Using generic linear systems theory, or by formulating detailed electromagnetic models, it can be shown that the recorded power can always be written in the form
\begin{align}
\label{B1}
P(t) & =  \int_{\cal D} {\rm d}^{3} {\bf r}_{1} \int_{\cal D} {\rm d}^{3} {\bf r}_{2} \int {\rm d} t_{1} \int {\rm d}t_{2} \, \overline{\overline{D}} (t;{\bf r}_{1},t_{1};{\bf r}_{2},t_{2}) \cdot \cdot \, \overline{\overline{E}} ({\bf r}_{1},t_{1};{\bf r}_{2},t_{2}),
\end{align}
where $\overline{\overline{D}} (t;{\bf r}_{1},t_{1};{\bf r}_{2},t_{2})$ is a dyadic field, called the {\em response tensor}, which characterises the energy-absorbing properties of the device. The spatial integrals are evaluated over the input reference surface of the device, whose outline defines some domain ${\cal D}$. $\overline{\overline{X}} \cdot \cdot \, \overline{\overline{Y}}$ denotes contraction of the dyads $\overline{\overline{X}}$ and $\overline{\overline{Y}}$ to a scalar, and corresponds to the `trace of the product' ${\rm Tr} \left[ X Y \right]$ when matrices are used to represent correlations between polarisations. $\overline{\overline{D}}$ and $ \overline{\overline{E}}$ can be transformed into the spatial ($k$) domain and/or the temporal frequency ($\omega$) domain, and the same functional form returns, suggesting that Equation (\ref{B1}) describes a basic physical process.

For common detectors, $\overline{\overline{D}} (t;{\bf r}_{1},t_{1};{\bf r}_{2},t_{2})$ is time-shift invariant, and if the field is statistically  stationary,  $\overline{\overline{E}} ({\bf r}_{1},t_{1};{\bf r}_{2},t_{2})$ is also time shift invariant. The detected power is then time invariant, and (\ref{B1}) reduces to the spectral form
\begin{align}
\label{B2}
P & = \int_{\cal D} {\rm d}^{3} {\bf r}_{1} \int_{\cal D} {\rm d}^{3} {\bf r}_{2} \int {\rm d} \omega \, \overline{\overline{D}} ({\bf r}_{1},{\bf r}_{2},\omega) \cdot \cdot \, \overline{\overline{E}} ({\bf r}_{1},{\bf r}_{2},\omega).
\end{align}
Mathematically, Equation (\ref{B2}) describes the full contraction of two tensor fields to a real-valued quantity, and is the most obvious way of creating a scalar, the measured power, from the second-order correlation function of the partially coherent field. In the abstract vector space of tensor fields, Equation (\ref{B2}) describes the orthogonal projection of a tensor that describes the state of coherence of the field onto a tensor that describes the state of coherence to which the detector is maximally receptive.

\begin{figure}
\noindent \centering
\includegraphics[trim = 4cm 20cm 7cm 4cm, width=60mm]{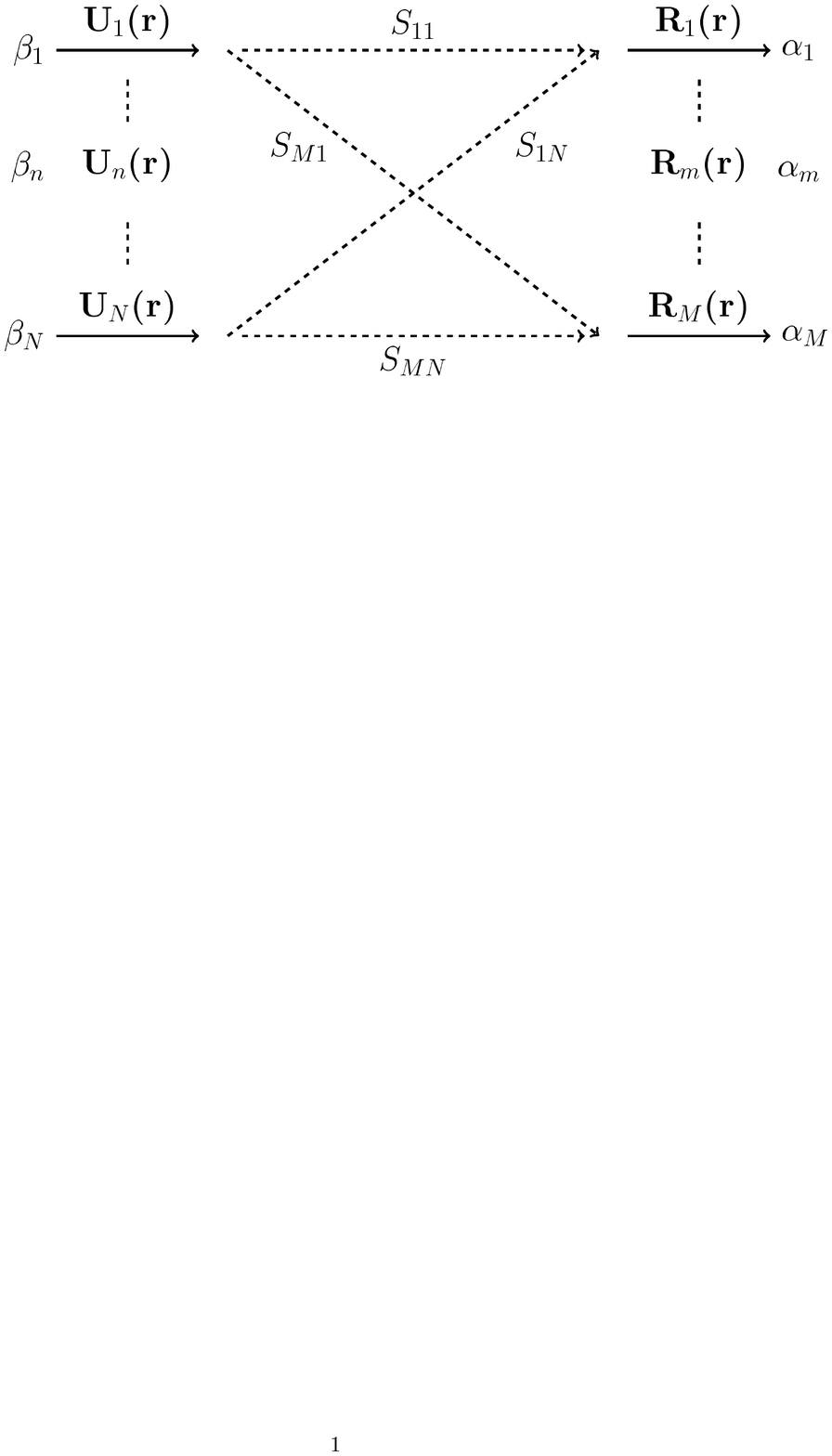}
\caption{Natural modes of the field ${\bf U}_{n} ({\bf r})$, carrying power $\beta_{n}$, couple to the natural modes of the detector ${\bf R}_{m} ({\bf r})$, having responsivity $\alpha_{m}$, through the scattering parameters $S_{mn}$. The total power absorbed, and recorded, is given by a sum of these processes.}
\label{figure1}
\end{figure}

Because $\overline{\overline{D}}$ and $ \overline{\overline{E}}$ are Hermitian, and square integrable, they can be diagonalised by the decompositions
\begin{align}
\label{B3}
\overline{\overline{D}} ({\bf r}_{1},{\bf r}_{2},\omega) & = \sum_{m} \alpha_{m} {\bf R}_{m} ({\bf r}_{1}) {\bf R}_{m}^{\ast} ({\bf r}_{2}) \\
\label{B4}
\overline{\overline{E}} ({\bf r}_{1},{\bf r}_{2},\omega) & = \sum_{n} \beta_{n} {\bf U}_{n} ({\bf r}_{1}) {\bf U}_{n}^{\ast} ({\bf r}_{2}),
\end{align}
where all quantities on the right are frequency dependent. The basis functions ${\bf R}_{m} ({\bf r})$ form an orthogonal set over ${\cal D}$, and likewise for ${\bf U}_{n} ({\bf r})$. Substituting Equations (\ref{B3}) and (\ref{B4}) in Equation (\ref{B2}) gives
\begin{align}
\label{B5}
P & = \int {\rm d} \omega \, \sum_{mn} \alpha_{m} (\omega) \beta_{n} (\omega) | S_{mn} (\omega) |^{2} \\ \nonumber
S_{mn} (\omega) & = \int_{\cal D} {\rm d} {\bf r} \, {\bf R}_{m}^{\ast} ({\bf r}) \cdot {\bf U}_{n} ({\bf r}).
\end{align}
which is called the {\em coupled mode model} \cite{Wth1,Sak1}. In Equation (\ref{B4}), the partially coherent field $\overline{\overline{E}}$ is described by an incoherent superposition of fully coherent fields ${\bf U}_{n} ({\bf r})$, each of which carries power $\beta_{n}$. In Equation (\ref{B3}), the response function $\overline{\overline{D}}$ is described by a set of complex-valued reception patterns ${\bf R}_{m} ({\bf r})$, each of which has some responsivity $\alpha_{m}$. The reception patterns are the individual degrees of freedom through which the device can absorb energy. In the $k$ domain, they correspond to the complex-valued angular beam patterns of the individual modes of the device. They are determined by the shape and size of the device (the boundary conditions), and the spatial coherence length of the solid-state processes responsible for absorption: such as electron-phonon interactions or spin-wave damping. According to Equation (\ref{B5}), the detected power is given by scattering, $S_{mn} (\omega)$, the natural modes of the field onto the natural modes of the detector: Figure \ref{figure1}. The coupling is maximised when the field modes and detector modes couple in one-to-one correspondence. From a photon perspective, one might now be concerned about the appearance of additional partition-noise effects.

Suppose that two different detectors, $a$ and $b$, are somewhere in an incoming optical beam: for example, two pixels in an imaging array. It can be shown \cite{Sak1}, using the Poisson mixture technique and Gaussian moment theorem, that the covariance $C[P_{a},P_{b}]$ of the outputs of the detectors due to fluctuations in the incident field are
\begin{align}
\label{B6}
& C[P_{a},P_{b}]  = \\ \nonumber
& \frac{1}{\tau} \int {\rm d} \omega \int_{{\cal D}_{a}} {\rm d}^{3} {\bf r}_{1}  \int_{{\cal D}_{a}} {\rm d}^{3}  {\bf r}_{2}
 \int_{{\cal D}_{b}} {\rm d}^{3} {\bf r}_{3}  \int_{{\cal D}_{b}} {\rm d}^{3} {\bf r}_{4} \,
\overline{\overline{D}}_{a} ({\bf r}_{1},{\bf r}_{2},\omega) \cdot \overline{\overline{E}} ({\bf r}_{2},{\bf r}_{3},\omega)
\cdot \cdot \, \overline{\overline{D}}_{b} ({\bf r}_{3},{\bf r}_{4},\omega) \cdot \overline{\overline{E}} ({\bf r}_{1},{\bf r}_{4},\omega) \\ \nonumber
& + \frac{\delta_{a,b}}{\tau} \int {\rm d} \omega  \int_{{\cal D}_{a}} {\rm d}^{3} {\bf r}_{1}  \int_{{\cal D}_{a}} {\rm d}^{3} {\bf r}_{2} \, \hbar \omega
 \overline{\overline{D}}_{a} ({\bf r}_{1},{\bf r}_{2},\omega) \cdot \cdot \, \overline{\overline{E}} ({\bf r}_{1},{\bf r}_{2},\omega).
\end{align}
The first term characterises the classical fluctuations, and the second term the photon shot noise. $a=b$ gives the variance in the output of a single detector. When $a \neq b$, the Dirac delta function $\delta_{a,b}$ indicates that photon absorption in different detectors is not correlated.

Equations (\ref{B2}) and (\ref{B6}) may seem complicated, but when $\overline{\overline{D}}$ and $ \overline{\overline{E}}$ are sampled spatially for numerical modelling they reduce to the Trace of a product of matrices. They are valuable when choosing the sizes, spacings and layouts of pixels in an imaging array to optimise efficiency and information recovery. For example, the modal approach is well suited to understanding straylight and radiation noise in pixels that couple poorly to the high-transmission modes of the preceding optical system. In addition, the response function technique can be used to model the behaviour of complete instruments, rather than just the detectors, leading to many applications.

\section{Power detection - a quantum perspective}

\label{sec_pow_qua}

Section \ref{sec_pow_det} describes a way of modelling power detectors, but if we adopt a quantum-mechanical approach, do we arrive at the same mathematical form? Consider an energy absorbing system having certain physical properties described by the Hermitian operators $\hat{A}$ and $\hat{B}$, and a source described by a generalised force $\hat{F}$, which may be an electric field, magnetic field, vector potential or some other perturbing quantity such as a strain field: Figure 2. If $\hat{H}_{\rm sys}$ and $\hat{H}_{\rm src}$ are the Hamiltonians of the system and source, the overall Hamiltonian is $\hat{H} =  \hat{H}_{\rm sys} + \hat{H}_{\rm src} + \hat{H}_{\rm int} =\hat{H}_{0} + \hat{H}_{\rm int}$, where
\begin{align}
\label{C1}
\hat{H}_{\rm int} (t) & = \kappa  \int_{\cal V} {\rm d}^{3} {\bf r} \,   \hat{A} ({\bf r},t) \cdot \hat{F} ({\bf r},t),
\end{align}
and $\kappa$ is a variational parameter. The interaction Hamiltonian $\hat{H}_{\rm int} (t)$ means that the source influences the time evolution of the system, and vice versa. If the force is constant over the volume of the system, $\hat{F} ({\bf r},t) = \hat{F} (t)$, it can be taken outside of the integral; if the force  is a scalar, $\hat{F} ({\bf r},t) = F ({\bf r},t)$, it merely scales the interaction energy. $\hat{F} ({\bf r},t)$ acts on the state space of the source, which in the case of electromagnetic radiation is the multi-mode Fock space of field. The connected property of the system $\hat{A} ({\bf r},t)$ acts on the multi-particle state space of the system, such as the electrons in a conductor. If the source and system are not entangled prior to the force being applied, the initial composite state is the tensor product $| \psi (t_{0}) \rangle = | \psi_{\rm sys} (t_{0}) \rangle | \psi_{\rm src} (t_{0}) \rangle$.

\begin{figure}
\noindent \centering
\includegraphics[trim = 4cm 20cm 7cm 4cm, width=60mm]{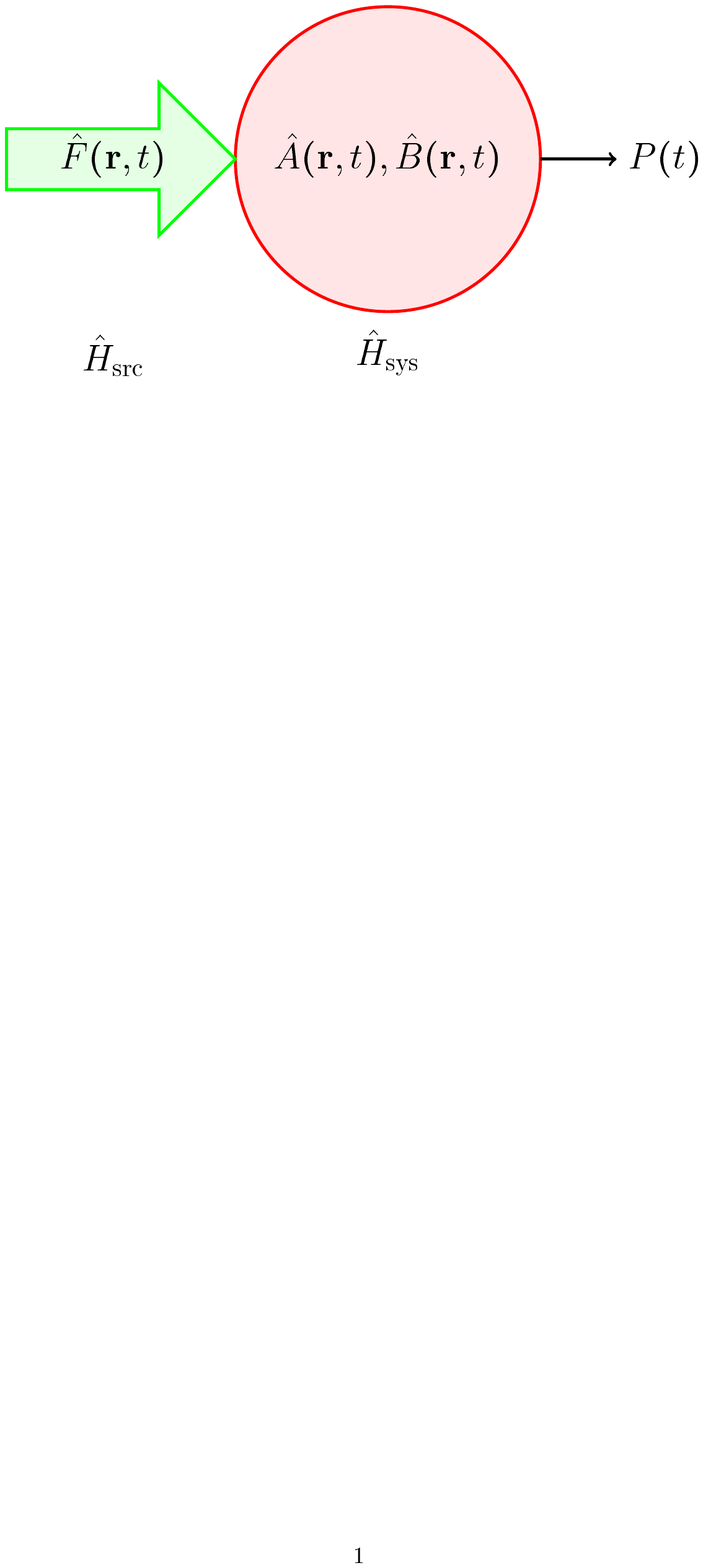}
\caption{Generalised force $\hat{F} ({\bf r},t)$ acts on an energy absorbing system having physical characteristics $\hat{A} ({\bf r},t)$ and $\hat{B} ({\bf r},t)$. The rate at which work is done on the system gives rise to a classical measure of instantaneous power $P(t)$.}
\label{figure2}
\end{figure}

In the Schr\"{o}dinger Picture, the composite state evolves from $t_{0}$ to $t$ according to the time evolution operator, $| \psi(t) \rangle = \hat{U}(t,t_{0}) | \psi(t_{0}) \rangle$. In the Heisenberg Picture, time evolution is attached to the operators themselves,  $\hat{A}^{\rm H}(t) = \hat{U}^{\dagger}(t,t_{0}) \hat{A}(t_{0}) \hat{U}(t,t_{0})$, leaving the states invariant $| \psi(t) \rangle = | \psi(t_{0}) \rangle$, and leading to the idea of measuring a quantity at some point in time. For perturbative influences, it is best to use the Interaction Picture, where part of the time evolution is attached to the states, and part to the Hermitian operators. Define a new time-shift operator $\hat{S}(t,t_{0}) = e^{+ i \hat{H}_{0} (t-t_{0}) /  \hbar} \hat{U}(t,t_{0})$, where the part that would have happened anyway in the absence of the perturbation is removed from $\hat{U}(t,t_{0})$ through time reversal,  $e^{+ i \hat{H}_{0} (t-t_{0}) /  \hbar}$. The states then evolve according to the perturbation, $| \psi(t)\rangle^{\rm I} =   \hat{S}(t,t_{0}) | \psi(t_{0}) \rangle$, and the operators according to the free evolution  $\hat{A}^{\rm I}(t) =  e^{+ i \hat{H}_{0} (t-t_{0}) /  \hbar} \hat{A}(t_{0}) e^{- i \hat{H}_{0} (t-t_{0}) /  \hbar}$. It can be shown that time shift operator, or scattering operator, is given by
\begin{align}
\label{C2}
\hat{S}(t,t_{0})  = \stackrel{\leftarrow}{\cal T} \left[ \exp \left\{ \left( \frac{-i}{\hbar} \right) \int_{t_{0}}^{t} {\rm d}t' \hat{H}^{I}_{\rm int} (t') \right\} \right] \hspace{10mm} t \ge t_{0},
\end{align}
where $\stackrel{\leftarrow}{\cal T}$ indicates that once the exponential is written as a power series, all operators  should be arranged in order of increasing time, from right to left. Although Equation (\ref{C2}) comes from the iterated solution of a differential equation, it can be appreciated, with some care, that if time is discretised, evolution occurs through a sequence of exponential factors where the interaction energy is approximately constant during each step. First-order theory only uses the first two terms of Equation (\ref{C2}),
\begin{align}
\label{C3}
 \hat{S} (t,t_{0}) & \approx 1 -  \frac{i}{\hbar} \int_{t_{0}}^{t}  {\rm d}t' \, \hat{H}^{I}_{\rm int} (t'),
\end{align}
linearising  $\hat{S} (t,t_{0})$ in $\kappa$. The higher-order terms, describing more complicated virtual processes, could be included if nonlinear behaviour is of interest.

Suppose that some other physical quantity $\hat{B} ({\bf r},t)$ responds to the perturbation. Without additional assumptions, straightforward algebra shows that
\begin{align}
\nonumber
\hat{B}^{H}  ({\bf r},t) & = \hat{S}^{\dagger} (t,t_{0}) \hat{B}^{I}  ({\bf r},t) \hat{S}(t,t_{0}) \\
\label{C4}
& \approx \hat{B}^{I}  ({\bf r},t) - \kappa \frac{i}{\hbar} \int_{t_{0}}^{t}  {\rm d}t' \, \int_{\cal V} {\rm d}^{3} {\bf r}' \,
\left[ \hat{B}^{I}  ({\bf r},t), \hat{A}^{\rm I} ({\bf r}',t') \right] \cdot \hat{F}^{I} ({\bf r}',t') \\ \nonumber
& \equiv \hat{B}_{0}^{\rm H} ({\bf r},t) + \kappa \Delta \hat{B}^{\rm H} ({\bf r},t),
\end{align}

The first term, $\hat{B}_{0}^{H} ({\bf r},t)$, describes the free evolution of the system, and the second term, $\Delta \hat{B}^{H} ({\bf r},t)$, describes the linearised change brought about by the perturbation. Being in the Heisenberg picture, the expectation value of $\hat{B}^{H}  ({\bf r},t)$ is found with respect to the state of the system at $t_{0}$, which is the reference time for the phase factors, and precedes the time at at which the perturbation turns on. Carrying out the same operation on the right of Equation (\ref{C4}):
\begin{align}
\label{C5}
\langle \Delta \hat{B}^{H}  ({\bf r},t) \rangle_{t_{0}} & = \frac{-i}{\hbar} \int_{-\infty}^{+\infty}  dt' \, \theta(t-t') \int_{\cal V} {\rm d}^{3} {\bf r}' \,  \langle \left[ \hat{B}^{I}  ({\bf r},t), \hat{A}^{\rm I} ({\bf r}',t') \right] \rangle_{t_{0}} \cdot \langle \hat{F} ({\bf r}',t') \rangle_{t_{0}},
\end{align}
which has factored because the source and system are not entangled at $t_{0}$. The upper limit on the integral has been changed by including the step function $\theta(t-t')$, which enforces causality, and the lower limit has been changed because the source is assumed to turn on after $t_{0}$. Equation (\ref{C5}) is  called Kubo's formula \cite{Kub1}, and is the quantum equivalent of a classical response function.

How should the expectation values be evaluated? The source and system are in definite quantum states at $t_{0}$, but because of the extremely large number of degrees of freedom involved (for example the numerous electrons in an absorbing film), we cannot hope to know, or wish to know, what they are! The expectation values are therefore calculated through $\langle \hat{X} \rangle = \sum_{n} P_{n} \langle n | \hat{X} | n \rangle = {\rm Tr} [ \hat{\rho} \hat{X} ]$ where $\hat{\rho} = \sum_{n} P_{n} | n \rangle \langle n |$, and $P_{n}$ is the probability that the state is one of a complete set of eigenstates, $| n \rangle$. The density operator $\hat{\rho}$ incorporates two uncertainties: (i) our lack of certainty about which state the configuration is in; (ii) nature's lack of certainty about which eigenstate the system will collapse into when a measurement is made. If the absorber is cooled by a refrigerator, the system's density operator is $\hat{\rho}_{\rm sys} = \exp [\hat{H}_{\rm sys}/kT_{\rm sys}]$; and if the source is thermal radiation emitted by a warm load $\hat{\rho}_{\rm src} = \exp[\hat{H}_{\rm src}/kT_{\rm src}]$. By introducing these thermodynamic operators, an {\it open quantum system} has been created, hiding the quantum mechanics of the refrigerator and thermal source from view.

Using density operators, and elevating the response to vector-valued quantities,
\begin{align}
\label{C6}
\langle \Delta \hat{B}^{H}  ({\bf r},t) \rangle_{t_{0}} & =  \int_{-\infty}^{+\infty}  {\rm d}t' \, \int_{\cal V} {\rm d}^{3} {\bf r}' \, \overline{\overline{D}}_{\rm BA}({\bf r},t; {\bf r}';t') \cdot {\rm Tr} \left[ \hat{\rho}_{\rm src} (t_{0}) \hat{F} ({\bf r}',t') \right] \\
\label{C7}
\overline{\overline{D}}_{BA}({\bf r}, t; {\bf r}',t') & = \frac{-i}{\hbar} \theta(t-t') {\rm Tr} \left[ \hat{\rho}_{\rm sys} (t_{0}) \left[ \hat{B}^{I}  ({\bf r},t), \hat{A}^{\rm I} ({\bf r}',t') \right] \right].
\end{align}
Equation (\ref{C7}) is the dyadic form of Kubo's formula \cite{Kub1}, which describes how macroscopic characteristics such as impedance, dielectric constant, and permeability emerge from the microscopic behaviour of the solid state system. The elements of $\overline{\overline{D}}(t,t')$ are called {\em retarded Green's functions}, and it can be shown that they describe rather beautifully how a system responds when an excitation is introduced. For example, the injection of an electron or hole at one space-time point may be correlated with the appearance of an electron or hole elsewhere. Crucially, there is no need to follow every degree of freedom, but simply to know how the system responds when excitations are introduced. Green's functions are used extensively for modelling the behaviour of materials, such as the bulk impedance of superconducting films \cite{Zub1} and proximity effects \cite{Bel1}.

To calculate the behaviour of a power detector, it is necessary to know the instantaneous rate at which work is done on the system by the source, which is given by
\begin{align}
\label{C8}
\hat{P} (t^{\prime \prime}) & =  \int_{\cal V} {\rm d}^{3} {\bf r} \,  \hat{F}( {\bf r}, t^{\prime \prime})  \frac{\rm d}{{\rm d}t^{\prime \prime}} \Delta \hat{A} ({\bf r},t^{\prime \prime}),
\end{align}
which should be compared with force times rate of change of displacement. Substituting $\Delta \hat{A} ({\bf r},t^{\prime \prime})$ and calculating the expectation value gives
\begin{align}
\label{C9}
\langle \hat{P} (t^{\prime \prime}) \rangle & =  \int_{\cal V}  {\rm d}^{3} {\bf r}   \int_{\cal V}  {\rm d}^{3} {\bf r}' \int_{-\infty}^{+\infty}  {\rm d}t' \, \,   \left\{ \frac{- i}{\hbar}  \frac{\rm d}{{\rm d}t^{\prime \prime}}  \theta(t^{\prime \prime}-t') {\rm Tr} \left[ \hat{\rho}_{\rm sys}( t_{0}) \left[  \hat{A}^{I}  ({\bf r},t^{\prime \prime}), \hat{A}^{I} ({\bf r}',t') \right] \right] \right\} \\ \nonumber
& \times {\rm Tr} \left[ \hat{\rho}_{\rm src} (t_{0}) \hat{F}( {\bf r}, t^{\prime \prime}) \hat{F}( {\bf r}', t') \right].
\end{align}

Equation (\ref{C9}) nearly has the form of Equation (\ref{B1}), but not quite: (i) Equation (\ref{C9}) describes energy flow into the system, but does not give the quantity that is recorded at the output. (ii) Equation (\ref{B1}) has surface integrals whereas Equation (\ref{C9}) has volume integrals, and so the system is a volumetric absorber, rather than having a reference surface. Strictly, the volume integrals need transforming into surface integrals, but if the source is uniform, the problem simplifies anyway. (iii) The source term corresponds to excitation in the absence of any scattering in the medium, and so ignores screening. If screening is included, the functional form of the coupled-mode model does not change,  but the expression for the response tensor does; say in the case of multi-layered patterned detector arrays \cite{Wth2}. (iv) Equation (\ref{C9}) is based on the instantaneous work done, and so potentially includes energy flowing in and out of the detector in a reactive manner. All of these items can be dealt with in a straightforward way, returning the functional form of (\ref{B1}).

For example in the case of (i), a detector has some responsivity, which converts the instantaneous power into the recorded output, such as a voltage, and this conversion is band limited. Equation (\ref{C9}) can be convolved with some causal response function $g(t-t^{\prime \prime})$, which describes the conversion process, to give
\begin{align}
\label{C10}
P(t) & =  \int_{\cal V} {\rm d} {\bf r}_{1} \int_{\cal V} {\rm d} {\bf r}_{2} \int {\rm d} t_{1} \int {\rm d}t_{2} \, K(t;{\bf r}_{1},t_{1};{\bf r}_{2},t_{2}) \, F({\bf r}_{1},t_{1};{\bf r}_{2},t_{2}),
\end{align}
where
\begin{align}
\label{C11}
K(t; {\bf r}_{1},t_{1} ; {\bf r}_{2},t_{2}) & = g(t-t_{1}) \frac{\rm d}{{\rm d}t_{1}} \left\{ \frac{- i}{\hbar} \Theta(t_{1}-t_{2}) {\rm Tr} \left[ \hat{\rho}_{\rm sys} (t_{0})  \left[  \hat{A}^{I}  ({\bf r}_{1},t_{1}), \hat{A}^{I} ({\bf r}_{2},t_{2}) \right] \right] \right\} \\
\label{C12}
F({\bf r}_{1}, t_{1}; {\bf r}_{2},t_{2}) & = {\rm Tr} \left[ \hat{\rho}_{\rm src} (t_{0}) \hat{F}( {\bf r}_{1}, t_{1}) \hat{F}( {\bf r}_{2}, t_{2}) \right].
\end{align}
Equation(\ref{C10}) is very similar to (\ref{B1}), but now the process responsible for energy absorption and the source fields are both described by quantum correlation functions, and so quantum properties are included.

Creating an output signal by simply smoothing the expected value of the absorbed power is somewhat arbitrary. An information-theoretic approach is as follows. Suppose that $P(u|v)$ is the conditional probability that an observer records output $u$ when the object being measured is in eigenstate $| v \rangle$. If the object is actually in state $\hat{\rho}_{\rm int}$, then
\begin{align}
\nonumber
P( u ) & = \int {\rm d} v \, P(u|v) \langle v | \hat{\rho}_{\rm int} | v \rangle \\
\label{C13}
& = {\rm Tr} \left[ \hat{W} (u) \hat{\rho}_{\rm int} \right],
\end{align}
where
\begin{align}
\label{C14}
\hat{W} (u) & = \int {\rm d} v \, P(u|v) | v \rangle \langle v |.
\end{align}
$\hat{W} (u)$ looks like a density operator, but is a {\em measurement operator}. Equation (\ref{C9}) can be cast in this way, where $| v \rangle$ is the state of the composite system, $| \psi(t) \rangle$, and $P(u|v)$ is closely related to $g(t-t^{\prime \prime})$. $\hat{W} (t)$ describes the act of acquiring classical information about the absorbing system, as it interacts quantum mechanically with the source. This shift of perspective is not merely pedantry; it is intimately related to the notion of back action, where the state of the object being probed changes as a consequence of information being accrued. In information theoretic descriptions of quantum measurement, the quantum system being probed (the source) first interacts with some other quantum system (the device), which may itself be warm and in a mixed state, and which then provides a classical estimator of some aspect of the source's behaviour.

According to Von Neumann, when a measurement is made, the system collapses onto the eigenstate of the eigenvalue recorded. An identical measurement, immediately after the first, returns the same result with certainty. But this projective approach to state collapse does not seem to fit with the language of imperfectly constrained and continuous measurement, which is needed in the case of electrical circuits and sensing. Measurement operators such as $\hat{W} (t)$ allow a more nuanced approach. If the object being probed is initially in some mixed quantum state $\hat{\rho}_{\rm int}$, then after measurement it `collapses' into some new mixed, as distinct from pure, quantum state $\hat{\rho}_{\rm fin}$. Subsequent identical measurements allow additional information to be gathered; rather than simply repeating the same result as if there is no information left to be extracted. Each time $\hat{W} (t)$ is applied, information is acquired, and the entropy falls. Understanding the relationship between quantum information theory and sensor physics is an important part of progressing quantum technology.

In the case of (iv), assume that the source is a single-mode, time-harmonic wave,
\begin{equation}
\label{C16}
\hat{F}( {\bf r}, t)  = f( {\bf r}) \hat{a}  e^{-i \omega_{0} t} + f^{\ast}( {\bf r}) \hat{a}^{\dagger}  e^{+i \omega_{0} t},
\end{equation}
where $f( {\bf r})$ is some general factor. Then
\begin{align}
\nonumber
& {\rm Tr} \left[ \hat{\rho}_{\rm src} (t_{0}) \hat{F}( {\bf r}_{1}, t_{1}) \hat{F}( {\bf r}_{2}, t_{2}) \right] \equiv \\ \label{C17}
&  f( {\bf r}_{1}) f( {\bf r}_{2}) e^{-i \omega_{0} (t_{1}+t_{2})} \langle  \hat{a}  \hat{a} \rangle +  f^{\ast}( {\bf r}_{1}) f^{\ast}( {\bf r}_{2}) e^{+i \omega_{0} (t_{1}+t_{2})} \langle  \hat{a}^{\dagger}  \hat{a}^{\dagger} \rangle \\ \nonumber
& + f( {\bf r}_{1}) f^{\ast}( {\bf r}_{2}) e^{-i \omega_{0} (t_{1}-t_{2})} \langle  \hat{a}  \hat{a}^{\dagger} \rangle +
f^{\ast}( {\bf r}_{1}) f( {\bf r}_{2}) e^{+i \omega_{0} (t_{1}-t_{2})} \langle \hat{a}^{\dagger}  \hat{a} \rangle
,
\end{align}

The first two terms, in $t_{1}+t_{2}$, are fast as $t_{1}$ and $t_{2}$ increase, and are removed by the output smoothing; the last two terms, in $t_{1}-t_{2}$, are slow, and result in the recorded power. If the source is in a coherent quantum state, having complex amplitude $a$, then $\langle  \hat{a}^{\dagger} \hat{a} \rangle \rightarrow a^{\ast} a$ and $\langle  \hat{a}  \hat{a}^{\dagger} \rangle \rightarrow a^{\ast} a + 1$, and for a high-occupancy state the source correlation function becomes a fully coherent classical correlation function. Therefore, the response function $K(t;{\bf r}_{1}, t_{1};{\bf r}_{2},t_{2})$  characterises the response to both classical and quantum sources; and its general properties can be imported from classical considerations, or by knowing how the device responds to quantum excitations.

Equation (\ref{B1}) is defined in terms of {\em average power absorbed}, but Equation (\ref{C10}) is based on instantaneous {\em work done}, which potentially includes energy flowing in and out of the device, say a thin-film absorber, in a reactive manner. Because the response function is time-shift invariant, it can be Fourier transformed in $t_{1} - t_{2}$: $K({\bf r}_{1}; {\bf r}_{2};  t_{1} - t_{2})  \mapsto K({\bf r}_{1}; {\bf r}_{2}; \omega)$. Describing (\ref{C10}) in the Fourier domain, and using the classical-coherent limit of Equation (\ref{C17}), it can be shown that $P(t) \propto \left[ 1 + \cos (2 \omega_{0} t) \right] \cos (\theta) + \sin(2 \omega_{0} t) \sin(\theta)$, where $\theta$ is the phase of $K({\bf r}_{1}; {\bf r}_{2}; \omega)$. For $-\pi/2 \le \theta \le + \pi/2$ the first term, which is proportional to $\cos(\theta)$, the {\em power factor}, is always positive and describes time varying power dissipated in the detector: the power that flows into and stays in the detector varies in time (a principle closely related to {\em homodyne} detection); it has a time-averaged value of unity (a principle exploited in power detectors). The second term, which is proportional to $\sin(\theta)$, has a time-averaged average value of zero, and describes energy sloshing in and out of the detector. Thus the real part of the response function, $\Re \left[ K({\bf r}; {\bf r}'; \omega) \right]$, characterises dissipation, and is a manifestation of Fermi's Golden Rule, and the imaginary part, $ \Im \left[ K({\bf r}; {\bf r}'; \omega) \right]$, energy storage. These processes happen at the input regardless of the smoothing action of the output filter, which  ensures that only the time average part of the dissipated power contributes to the recorded output. One word of warning is that in physics and engineering, the roles of the real and imaginary parts of $K({\bf r}_{1}; {\bf r}_{2}; \omega)$ are swapped. In engineering, it is the real part of an impedance that describes loss, $R + j \omega L$, but in physics, it is the imaginary part of Kubo's susceptance that describes loss, and radiates noise as described by the {\em dissipation fluctuation theory} \cite{Flc1}. This difference can be traced to Equation (\ref{C8}), where physicists use the dynamical variables  $\hat{F}( {\bf r}, t)$ and $\Delta \hat{A} ({\bf r},t)$, whereas engineers use $\hat{F}( {\bf r}, t)$ and ${ \rm d} \Delta \hat{A} ({\bf r},t) / {\rm d} t$, and so is a matter of convention only. Finally, we note that for vector-valued fields, and adopting the engineering convention, it is the Hermitian part of the response tensor that corresponds to dissipated energy, and the anti-Hermitian part that corresponds to stored energy. Thus $\overline{\overline{D}}({\bf r}_{1}; {\bf r}_{2}; \omega)$ is the Hermitian part of $\overline{\overline{K}}({\bf r}_{1}; {\bf r}_{2}; \omega)$ when Equation (\ref{C10}) is upgraded to its full tensor form, and used in Equation (\ref{B1}).

The coupled mode model, Section \ref{sec_mul_det}, is based on the fact that the response function and correlation function in Equation (\ref{B1}) are both Hermitian and so can be diagonalised. In the quantised case, the response function is the same as the classical case, and so it can be diagonlised as before giving the modes of the detector. The field correlation function, however, is not Hermitian, Equation (\ref{C17}): the positive and negative parts have different physical interpretations. The positive frequency part describes photon absorption, whereas the negative frequency part describes photon emission, including spontaneous emission. This difference leads to characteristic features of thermal fields. For a more complete description of the quantised case, the response function and quantum correlation function should be split into their Hermitian and anti-Hermitian parts. The behaviour becomes more involved, but the overall scheme still describes the way in which the degrees of freedom present in the source couple to the degrees of freedom in the system available for absorbing energy. One effect is that a detector can radiate energy back into the source, and this radiated energy, which is fluctuating, can act as a noise source because it causes the energy stored in the device to vary: fluctuating power travels in both directions. Normally, noise is considered to be something that comes in from outside! If the device is at a higher temperature than the source, including internal heating such as hot-electron effects, the radiation noise can be greater than the noise associated with the source itself. Once again it is clear that the source and detector must be considered parts of a collected whole if all aspects of behaviour are to be understood.

The fact that $\overline{\overline{D}}({\bf r}_{1}; {\bf r}_{2}; \omega)$ characterises power absorption when classical sources are used is of considerable importance, because it shows that the complex-valued reception patterns, or antenna patterns, ${\bf R}_{m} ({\bf r}_{1})$ and the associated responsivities $\alpha_{m}$ can be determined interferometrically through power measurements alone. The technique is called Energy Absorption Interferometry (EAI) \cite{EAI1} and is closely related to aperture synthesis astronomy, but now the device under test is radiated with two phase locked coherent sources, rather than measuring the angular correlations emitted by thermal sources. Moreover, $\overline{\overline{D}}({\bf r}_{1}; {\bf r}_{2}; \omega)$ connects principles of reciprocity, where the dynamical degrees of freedom responsible for absorbing energy are the same as the degrees of freedom responsible for imposing near-field and far-field correlations on the thermally radiated fields \cite{Rec1}.

\section{Linear amplifiers - a classical perspective}

\label{sec_amp_cla}

Many aspects of sensing relate to measuring voltages and currents, or at least to amplifying weak signals to a level where classical signal processing can be performed. Rather than analysing complex circuits, it is common to place standard configurations in black boxes, to drive the ports with a set of independent variables (voltage and/or current) and to observe the response through a set of dependent variables. This approach focuses on those degrees of freedom accessible from the outside, and gives rise to small-signal impedance, admittance and hybrid circuit parameters. The multiport network approach leads to many general theorems, and can answer questions such as `is this device capable of producing gain' and if so `what embedding network is needed to achieve it'. In addition, voltage and current sources can be added to the ports to represent internally generated noise. There must be one noise source for every dependent variable, and every pair of noise sources has a complex correlation coefficient. The noise sources can be referenced to other ports for convenience. Additionally, external circuit elements may also produce noise.

In the case of amplifiers, a two-port network is sufficient. The noise sources are usually referenced to the input port, and take the form of a parallel current source and series voltage source \cite{Two1}. The correlation coefficient between them can be represented by introducing a fictitious noise impedance. Given an amplifying device, in a black box, one usually wishes to achieve three things: (i) Maximise the power gain by ensuring that the signal-source impedance $Z_{\rm s, pow}$ is conjugately matched to the impedance seen at the input of the loaded device. (ii) Choose a source impedance that interferes, and decorrelates to the highest possible degree, the currents induced in the output load by the two noise sources, as this minimises the recorded noise. There is some particular signal-source impedance $Z_{\rm s, nse}$ that minimises an amplifier's overall noise: a process called {\em noise matching}. (iii) Ensure that these optimisations do not result in the active device oscillating. Usually $Z_{\rm s, pow} \ne Z_{\rm s, nse}$ and ingenious schemes are needed to align the impedances, or simply to select the best compromise. The matter of whether it is best to have high gain or low noise is quantified through the concept of {\em noise measure}.

At high frequencies, when the wavelength is smaller than the dimensions of the circuit, it is beneficial to use a travelling wave representation. The ports of the black box are loaded with transmission lines having real characteristic impedance $Z_{0}$. The independent variables are the amplitudes $a(t)$ of the waves incident on the ports; and the dependent variables are the amplitudes $b(t)$ of the waves travelling away from the ports \cite{Kur1}. The relationships, in the frequency domain, between the  voltage $v(\omega)$ and current $i(\omega)$ at some reference plane and the complex wave amplitudes are
\begin{align}
\label{D1}
a(\omega) & = \frac{1}{2 \sqrt{Z_{0}}} \left[ v(\omega) + i(\omega) Z_{0} \right] \\ \nonumber
b(\omega) & = \frac{1}{2 \sqrt{Z_{0}}} \left[ v(\omega) - i(\omega) Z_{0} \right].
\end{align}
Generally, $a(\omega)$ and $b(\omega)$ are stochastic quantities, which must be averaged over an ensemble.  The average power spectral density incident on a port is $S_{\rm a} = E \left[ a a^{\ast} \right]$. Internal noise is represented by allowing noise waves to travel away from the ports even in the absence of external excitation. As in the discrete case, the travelling wave amplitudes may be correlated, and so a correlation matrix is needed whose diagonal elements are spectral powers.

\begin{figure}
\noindent \centering
\includegraphics[trim = 4cm 14cm 7cm 6cm, width=70mm]{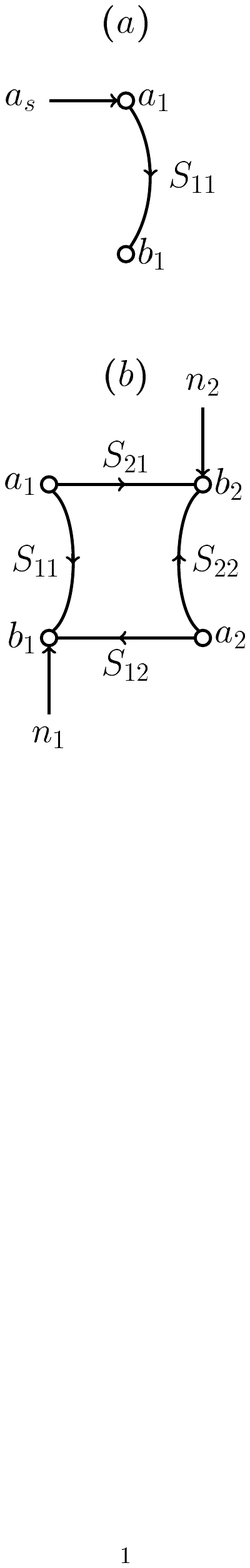}
\caption{Scattering parameter representations: (a) One-port network with a signal source $a_{s}$. (b) Two-port network with internal noise sources $n_{1}$ and $n_{2}$.}
\label{figure3}
\end{figure}

A multiport device is then represented by a signal flow graph. Figure \ref{figure3} shows signal flow graphs of (a) a one-port network with source, and (b) a two port network with internal noise sources. Figure 4 (a) shows the signal flow graph of a two-port network with the internal noise sources {\em referenced} to the input. Often, two-port networks are cascaded, and it is necessary to know the complex amplitude of the wave travelling away from the output in terms of the wave incident on the input. The existence of loops in connected networks, creating internal resonances, makes the analysis of signal flow graphs awkward. In the 1950's Mason introduced a {\em non-touching loop rule} \cite{Msn1} that allows expressions for interdependencies to be derived (Transducer Gain, Available Gain, Maximum Stable Gain, reflection coefficients, etc.)

\begin{figure}
\noindent \centering
\includegraphics[trim = 4cm 12cm 6cm 6cm, width=70mm]{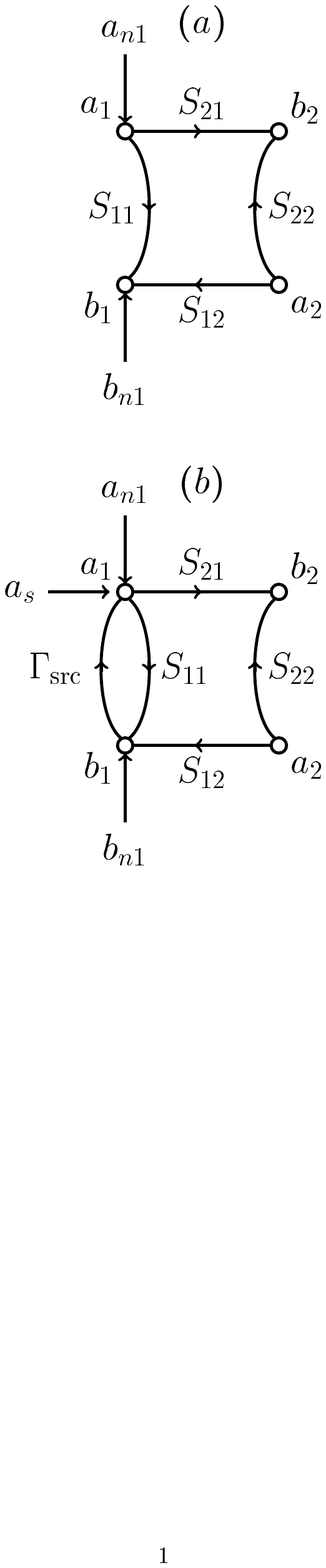}
\caption{(a) Two-port scattering parameters with the internal noise sources referenced to the input. (b) Complete amplifier with signal source reflection coefficient $\Gamma_{\rm src}$.}
\label{figure4}
\end{figure}

Consider the signal flow graph shown in Figure \ref{figure4} (a). The amplifying device is represented by a two-port network, with its internal noise sources referenced to the input. The sources comprise a noise wave effectively incident on the input $a_{n1}$, a noise wave travelling away from the input $b_{n1}$, and a complex correlation coefficient $\Gamma$ between them. As discussed in Section \ref{sec_rad_nse}, a perfectly matched passive warm termination radiates power into a single mode transmission line. Therefore, it is possible to describe power spectral densities in terms of equivalent temperatures: $T_{\rm a} = E \left[ a a^{\ast} \right] / k$ and $T_{\rm b} = E \left[ b b^{\ast} \right] / k$, where it is conventional to use the Rayleigh-Jeans limit in the definition: and it is only a definition. Also, the complex correlation coefficient between the wave amplitudes, $\Gamma$,  can be written as a complex-valued  `temperature' $T_{\rm c} = \Gamma \sqrt{ T_{\rm a} T_{\rm b}}$.

If the device is connected to noiseless terminations having impedance $Z_{0}$, the noise power effectively incident on the device $E \left[ a a^{\ast} \right]$ accounts for all of the noise appearing at the output, and the noise temperature becomes $T_{\rm n} =  T_{\rm a}$. However, a noise wave also travels away from the input, having noise temperature $T_{\rm b}$, and there is no reason why this should not be significantly larger than $T_{\rm a}$. Commercial suppliers report the noise temperature $T_{\rm a}$, but do not report $T_{\rm b}$, and so care is needed because noise power is fed back into the source. If the supplier has done a good job of minimising the external effects of the internal sources, $T_{\rm c} = 0$: otherwise you could find a source impedance that reduces the noise temperature below that claimed by the manufacturer!

If an amplifier is connected to a source having a non-zero reflection coefficient $\Gamma_{\rm src}$, referenced to $Z_{0}$, Figure \ref{figure4} (b), the outward travelling wave is partially reflected back in, and if $b$ is correlated with $a$, the noise wave travelling away from the output can, because of constructive interference, be significantly enhanced. The noise temperature is given by
\begin{align}
\label{D2}
T_{\rm n} = T_{\rm a} + |\Gamma_{\rm src}|^{2} T_{\rm b} + 2 T_{\rm c} |\Gamma_{\rm src}| \cos(\phi_{\rm c} + \phi_{\rm src}).
\end{align}

If the phase of the source reflection coefficient, $\phi_{\rm src}$, changes rapidly with frequency, say due to a long interconnecting cable, the noise temperature can vary widely and rapidly, with a peak variation $T_{\rm pk} = 4 T_{\rm c} | \Gamma_{\rm src} |$. Equation (\ref{D2}) does not include the input reflection coefficient of the amplifier, and so this equality holds regardless of whether the input of the amplifier is matched or not. Commercial amplifiers, as distinct from transistors, are noise matched internally, meaning that $T_{\rm c}=0$ at band centre. It can also be shown that when the source reflection coefficient $\Gamma_{\rm src}$  and input reflection coefficient of the amplifier $\Gamma_{\rm amp}$ are non zero, a resonant noise wave exists on the input transmission line, scaling as $\propto 1/|1-\Gamma_{\rm src} \Gamma_{\rm amp}|^{2}$, which can be large when $\Gamma_{\rm src} = \Gamma_{\rm amp}^{\ast}$. Here $\Gamma_{\rm amp} = S_{11}$. Another hazard is that most amplifiers have a large forward gain $|S_{21}|^{2}$ and tiny reverse gain $|S_{12}|^{2}$. Some ultra-low-noise amplifiers, such as certain parametric amplifiers, can have $|S_{12}|^{2} \approx 1$, and even gain in the reverse direction. Forward travelling noise from a second stage can be brought forward to the input of the first stage, and increase the overall noise temperature. To make sure that the system noise temperature is insensitive the source reflection coefficient, it is prudent to place a cooled circulator in front of the first amplifier. The noise-isolating role of the circulator has nothing to do with matching the input for maximum power gain, and because circulators are lossy and cumbersome, they are usually considered to be a technological nuisance.

Before closing this section, it is beneficial to reconsider flow graphs. In technical literature, analysis is carried out using Mason's rule, and for cascaded networks only having sources at the input and output ports this is sufficient. In the case of complicated networks, having internal noise sources, it is algebraically tedious to trace the effect of every source to the external ports. To make matters worse, the internal sources may be correlated and only the correlation matrix of the internal sources is known. Then it is only possible to calculate, even in principle, the correlation matrix of the noise waves appearing at the external ports. In the quantum case, one only has correlation functions of the kind $\langle \hat{b}_{{\rm s},i} \hat{b}_{{\rm s},j} \rangle$, and so a direct mapping from the source correlation functions to output correlation functions is needed.

To this end, the {\em connection matrix method}, which is built on the rich mathematical topic of {\em directed flow graphs}, is valuable. For any general network having $M$ nodes, and with $N$ sources entering the nodes, collect the travelling wave amplitudes at the nodes into column vector ${\mathsf d}$, and the travelling wave amplitudes of the sources entering the nodes into column vector ${\mathsf n}$, then
\begin{align}
\label{D3}
{\mathsf d} & =  {\mathsf C} {\mathsf d} + {\mathsf n} \\ \nonumber
& = \left[ {\mathsf I} - {\mathsf C} \right]^{-1} {\mathsf n} \\ \nonumber
& = {\mathsf K} {\mathsf n},
\end{align}
where the $i,j$th entry in ${\mathsf C}$ is the complex scattering parameter that connects node $j$ to node $i$. $N$ may be smaller than $M$, resulting in ${\mathsf n}$ having some zero entries. ${\mathsf I}-{\mathsf C}$ is sparse because only a small number of nodes are connected directly. Only the correlation matrix of the sources is known, ${\mathsf N}_{\rm s}$, and so it is only possible to calculate the correlation matrix of the resulting travelling waves, ${\mathsf N}_{\rm c}$. Equation (\ref{D3}) gives
\begin{align}
\label{D4}
{\mathsf N}_{\rm c} & =  {\mathsf K} {\mathsf N}_{\rm s} {\mathsf K}^{\dagger},
\end{align}
from which the correlation matrix of the waves of interest can be extracted. To carry out calculations when quantum noise is present, it is necessary to be careful about vacuum-state noise, because vacuum noise enters through even seemingly unused ports. The bosonic commutation relationships between the travelling wave amplitudes on the port transmission lines must be maintained.

\section{Linear amplifiers - a quantum perspective}

\label{sec_amp_qua}

\subsection{Quantum equivalent circuits}

\label{sec_eqv_qua}

\begin{figure}
\noindent \centering
\includegraphics[trim = 4cm 19cm 7cm 4cm, width=70mm]{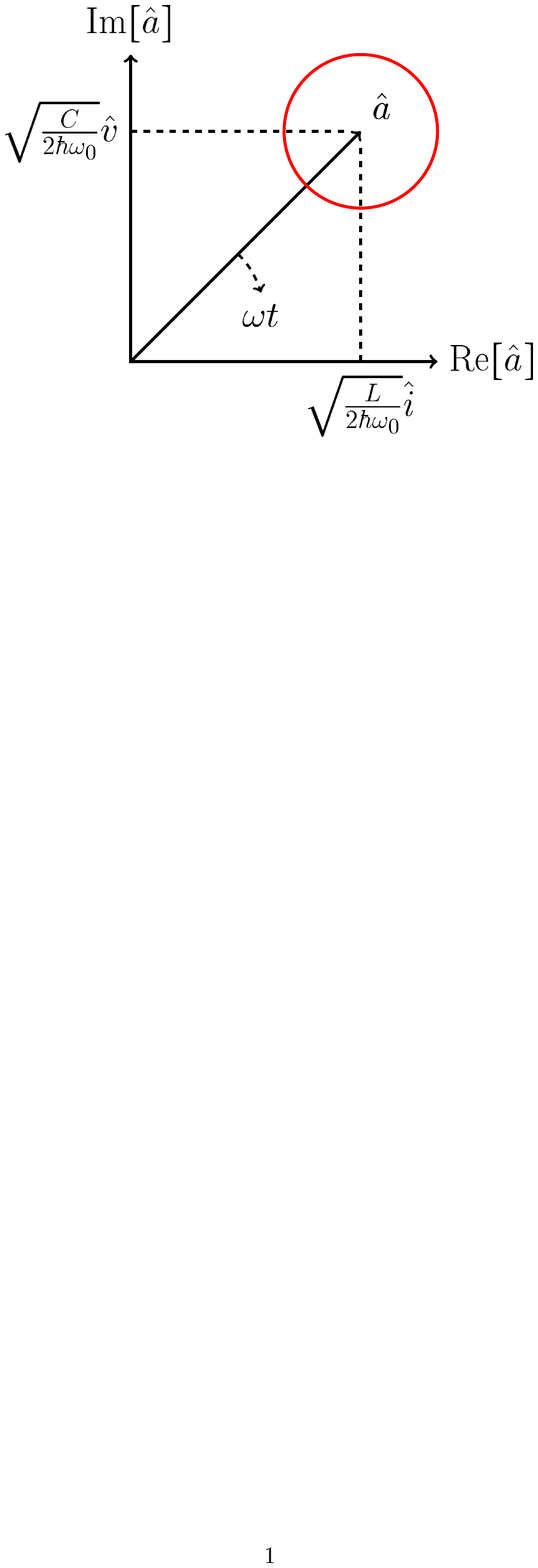}
\caption{Annihilation operator as a phasor in the complex plane. The real and imaginary parts are the current and voltage respectively. The red circle shows the variance on measurements.}
\label{figure5}
\end{figure}

At low frequencies simple electrical circuits are modelled using discrete components, which are then quantised through Lagrangian methods \cite{Lng1}, but in the case of complicated circuits, one is left searching for the Lagrangian that gives the correct answer! Consider an $L$-$C$ resonator made of discrete components. It can be shown that a perfect classical voltage or current source places the resonator in a coherent state $|a\rangle$. $\hat{a}$ is not however Hermitian, and the complex amplitude $a$ is not directly measurable: it has two degrees of freedom, amplitude and phase. Instead, we are left with three possible real-valued measurements: the energy in the mode, the voltage across the resonator, and the current through the resonator. The voltage and current operators are
\begin{align}
\label{E1}
\hat{v} (t) = i \left( \frac{\hbar \omega_{0}}{2 C} \right)^{1/2} \left( \hat{a}^{\dagger} e^{i \omega_{0} t} - \hat{a} e^{-i \omega_{0} t} \right) \\ \nonumber
\hat{i} (t) = \left( \frac{\hbar \omega_{0}}{2 L} \right)^{1/2} \left( \hat{a}^{\dagger} e^{i \omega_{0} t} + \hat{a} e^{-i \omega_{0} t} \right).
\end{align}
If the resonator is in a coherent state, the number of excitations $n \equiv |a|^{2}$ is Poisson distributed. If one of the quadrature components, either $v$ or $i$, is measured repeatedly, with the system in the same coherent state before every measurement is made, the distributions show $(\Delta v)^{2} =  \hbar \omega_{0} / 2 C$ and $(\Delta i)^{2} = \hbar \omega_{0} / 2 L$; these uncertainties do not depend on occupancy, and are minimum uncertainty states. This behaviour is illustrated in Figure \ref{figure5}. Now, however, quantum mechanics throws up a issue because $[i,v] = i \hbar  \omega_{0}^{2}$, and so according to the generalised uncertainty relationship $ \Delta A  \, \Delta B  \ge \left| \langle i \left[ \hat{A},\hat{B} \right] \rangle \right| / 2$, it follows that  $\Delta v \Delta i \ge \hbar \omega_{0}^{2} /2$. If the voltage is measured with an accuracy greater than the intrinsic uncertainty, $\sqrt{\hbar \omega_{0} / 2 C}$, then a subsequent measurement of current without re-establishing the state gives a variation that is considerably larger than $\sqrt{\hbar \omega_{0} / 2 L}$, and vica versa. It seems that immediate sequential measurements of voltage and current must be constrained by Heisenberg's uncertainty principle, in the same way that measurements of position and momentum are constrained in a freely oscillating mechanical resonator. A measurement of voltage or current leads to a {\em backaction} that changes the distribution that must be used on subsequent measurements. This general reasoning neglects to the role of dissipation, which leads to a finite quality factor $Q$, and introduces a time scale over which excitations are lost to the heat bath. In some cases with carefully chosen apparatus, it is possible to extract information without causing the state to change: a technique called {\em Quantum Non-Demolition} measurement.

If a resonator is weakly coupled to a heat bath having temperature $T_{\rm p}$, expectation values can be calculated using the density operator $\hat{\rho}_{\rm thm}$, giving
\begin{align}
\label{E2}
(\Delta v)^{2} & =  \frac{\hbar \omega_{0}}{2C} \left[  2 \langle n \rangle + 1 \right]  =  \frac{\hbar \omega_{0}}{2C}   \coth (\hbar \omega_{0} / 2kT_{\rm p}) \\ \nonumber
(\Delta i)^{2} & =  \frac{\hbar \omega_{0}}{2L} \left[  2 \langle n \rangle + 1 \right]  =  \frac{\hbar \omega_{0}}{2L}   \coth (\hbar \omega_{0} / 2kT_{\rm p}).
\end{align}
Weak coupling implies a low-value series resistor or a high-value parallel resistor, but Equation (\ref{E2}) has been derived without analysing the behaviour of an R-L-C circuit---only the density operator was used. In classical circuit analysis, a spectral noise voltage source, $(\Delta v)^{2} = 4 k T_{\rm p} R$, must be included in series with any resistor; in the quantum case, this expression changes to $(\Delta v)^{2} = 2 \hbar \omega_{0} R  \coth (\hbar \omega_{0} / 2kT_{\rm p})$. When circuit analysis is carried out, these spectral quantities are, effectively, multiplied by a bandwidth to give the actual variance in voltage.
For a single-pole resonator, which has a Lorentzian frequency response, the bandwidth is $ \pi f_{0} / 2Q = \pi f_{0} \omega_{0} L / 2 R$, and Equation (\ref{E2}) is recovered. In the limit $\hbar \omega_{0} / T_{\rm p} \rightarrow 0$, the classical expression $(\Delta v)^{2} \rightarrow  4kT_{\rm p}R$ is found. The transition from classical to quantum behaviour occurs when $\hbar \omega_{0} \approx kT_{\rm p}$, which for a 5 GHz resonator happens at $T_{\rm p}=$ 240 mK. Dilution refrigerator technology routinely achieves 10 mK, showing that even RF circuits can be operated in the quantum regime, motivating the need for quantum circuit theory at low temperatures.

The density operator $\hat{\rho}_{\rm thm}$ is based on thermodynamic considerations, and so tacitly assumes that the mechanism responsible for dissipation behaves as a weakly coupled heat bath having a large number of degrees of freedom. The overall quantum system is no longer closed, because no attempt is being made to track the behaviour of every degree of freedom: such as the electron-phonon system in a resistor. Strictly, the density operator no longer obeys von Neumann's differential equation for quantum operators, and must be described by the more complicated dynamics of the {\em Lindblad Master Equation}, which
unlike Schr\"{o}dinger's equation, describes the time evolution of mixed quantum states.

In the case of high-frequency circuits, a transmission-line representation with scattering parameters is usually best. Transmission lines are easily quantised, but with a note of caution. The voltage and current at every point are in phase, and related through the real-valued characteristic impedance $Z_{0}$. The voltage and current commute, and so are not restricted by Heisenberg's uncertainty relationship. In fact, if the instantaneous voltage is measured, the current is already known through $ V / Z_{0}$. In the case of transmission lines, voltage and current are compatible observables, whereas
the voltage and its time derivative, or the current and its time derivative, the {\em quadrature components}, are not.

When combining quantised transmission line theory with scattering parameter representations, there is an unfortunate clash of notation. In classical scattering parameter theory, $a(\omega) = v^{+} (\omega) / \sqrt{Z_{0}}$ and  $b (\omega) = v^{-} (\omega) / \sqrt{Z_{0}}$ are the normalised complex amplitudes of the counter-propagating waves $v^{+} (\omega)$ and $v^{-} (\omega)$, such that $|a(\omega)|^{2}$ and $|b(\omega)|^{2}$ are power spectral densities. But each wave has a creation and annihilation operator, and so we introduce the operator pairs $(\hat{a}(\omega),\hat{a}^{\dagger}(\omega))$ and
$( \hat{b}(\omega),\hat{b}^{\dagger}(\omega))$ for the forward and backward travelling waves respectively. After quantisation, Equation (\ref{D1}) becomes
\begin{align}
\label{E3}
\frac{\hat{v}^{+}(\omega)}{\sqrt{Z_{0}}} & = \frac{1}{2 \sqrt{Z_{0}}} \left[ \hat{v}(\omega) + \hat{i}(\omega) Z_{0} \right]= \left[ \frac{\hbar \omega}{2} \right]^{1/2}  \hat{a}  (\omega) \\ \nonumber
\frac{\hat{v}^{-}(\omega)}{\sqrt{Z_{0}}} & = \frac{1}{2 \sqrt{Z_{0}}} \left[ \hat{v}(\omega) - \hat{i}(\omega) Z_{0} \right] = \left[ \frac{\hbar \omega}{2} \right]^{1/2}  \hat{b}  (\omega),
\end{align}
where $\hat{v}(\omega)$ and $\hat{i}(\omega)$ are the voltage and current at a plane, say the port of a network. The average one-sided power spectral density flowing in the forward direction is given by the symmetrised form $\hat{s}^{+} (\omega) = ( \hbar \omega / 2 )  \left( \hat{a}  (\omega)  \hat{a}^{\dagger}  (\omega) +  \hat{a}^{\dagger}  (\omega)\hat{a}  (\omega) \right) = ( \hbar \omega / 2 )  \left\{ \hat{a}  (\omega) , \hat{a}^{\dagger}  (\omega)  \right\}$, and similarly for the reverse direction.

In a travelling-wave representation, the annihilation operators of the outgoing waves depend linearly on the annihilation operators of the incoming waves, with the constants of proportionality being the scattering parameters. This view is plausible because if the incoming waves are in high-occupancy coherent states, the outputs must correspond to those of a classically driven system. For a multiport network, the input operators act on the tensor product of the `input states': $| p_{1} \rangle \cdots | p_{m} \rangle  \cdots | p_{M} \rangle$. The output operators therefore act on the same state space: the outcomes of measurements
on the outgoing waves are are described in terms of the states of the incoming waves. The scattering parameters are essentially complex probability amplitudes. In general terms, the waves incident on the ports do not have have to be in coherent states, and may even be in mixed states, such as thermal states, but in order for the scheme to be self consistent, the vacuum states of seemingly undriven ports must be included.

It is often the case that incoming radiation is described solely in terms of  quantum correlation functions (for example $\langle \hat{a} (\omega) \hat{a}^{\dagger}(\omega) \rangle$), and then only outgoing correlation functions can be determined. With care relating to ports in the vacuum state, this mapping can be achieved through the connection matrix method, Equation (\ref{D4}).

Consider a multiport network, where one port comprises the input, another the output, and where $M > 2$; in other words, a microwave two-port network connects internally to a set of `internal' ports that influence the output but whose states are never measured.  One way of eliminating interest would be to take the partial trace over the `internal ports', to yield the two-port behaviour. Another approach is to say that
\begin{align}
\label{E4}
\left(
  \begin{array}{c}
    \hat{b}_{1} \\
    \hat{b}_{2} \\
  \end{array}
\right) & =
\left(
  \begin{array}{cc}
    S_{11} & S_{12} \\
    S_{21} & S_{22} \\
  \end{array}
\right)
\left(
  \begin{array}{c}
    \hat{a}_{1} \\
    \hat{a}_{2} \\
  \end{array}
\right) +
\left(
  \begin{array}{c}
    \hat{n}_{1} \\
    \hat{n}_{2} \\
  \end{array}
\right), \\ \nonumber
\hat{\mathsf b} & = {\mathsf S} \hat{\mathsf a} + \hat{\mathsf n},
\end{align}
where for brevity explicit reference to $\omega$ is dropped. The vector-valued operator $\hat{\mathsf n}$ contains linear contributions from the internal ports, and acts on a suitably extended state space. One may be tempted to ignore vacuum contributions from the internal ports, but if this is done, the output operators do not then satisfy bosonic commutation relationships. It is now clear that even vacuum states are likely to contribute to the output in the form of an additive `noise' term. In many cases, this noise will be thermal, and in some cases may be at an effective temperature higher than the physical temperature of the device.

\subsection{Quantum noise temperature}

\label{sec_qua_nse_tmp}

Even at low physical temperatures $T_{\rm p} \ll \hbar \omega / k $, and without internal heating such as hot electron effects, `noise' appears at the output in terms of a weighted linear combination of vacuum states, because of $\hat{\mathsf n}$. Therefore, every device must have some minimum noise temperature. What is the minimum noise temperature of a multiport network? This question depends on the properties of ${\mathsf S}$, such as reciprocity, unitarity, and even linearity, but in the case of an ideal two-port amplifier, with $S_{11}=S_{22}=0$, a simple but compelling argument is as follows. Equation (\ref{E4}) gives $\hat{b}_{2} = S_{21} \hat{a}_{1} + \hat{n}_{2}$, and because the source and noise terms correspond to different degrees of freedom,
\begin{align}
\label{E5}
[\hat{b}_{2},\hat{b}^{\dagger}_{2}] & = |S_{21}|^{2} [\hat{a}_{1},\hat{a}^{\dagger}_{1}] + [\hat{n}_{2},\hat{n}^{\dagger}_{2}] \\ \nonumber
[\hat{n}_{2},\hat{n}^{\dagger}_{2}] & = 1 - |S_{21}|^{2},
\end{align}
where the second line follows because the operators correspond to travelling waves on transmission lines: $[\hat{a}_{1},\hat{a}^{\dagger}_{1}] = 1$ and $[\hat{b}_{1},\hat{b}^{\dagger}_{1}]=1$. The one-sided spectral density of the power travelling away from the output is given by
\begin{align}
\label{E6}
s^{b} (\omega) = |S_{21}|^{2} s^{a} (\omega) + s^{n} (\omega).
\end{align}
$|S_{21}|^{2}$ appears as the transducer power gain of the amplifier, as in the classical case.

For any operator, $\hat{X} \hat{X}^{\dagger} = \left[ \hat{X}, \hat{X}^{\dagger} \right] / 2 + \left\{ \hat{X}, \hat{X}^{\dagger} \right\}/2$, and so  $\langle \left\{ \hat{n}_{2}, \hat{n}^{\dagger}_{2} \right\} \rangle  = 2 \langle \hat{n}_{2} \hat{n}^{\dagger}_{2}\rangle - \langle \left[ \hat{n}_{2}, \hat{n}^{\dagger}_{2} \right] \rangle$. For an amplifier having power gain,$|s_{21}|^{2} > 1$,
\begin{align}
\label{E7}
s^{n} (\omega) =  \frac{\hbar \omega}{2} \langle \left\{ \hat{n}_{2}, \hat{n}^{\dagger}_{2} \right\} \rangle & \ge - \frac{\hbar \omega}{2} \langle  \left[ \hat{n}_{2}, \hat{n}^{\dagger}_{2} \right] \rangle = \frac{\hbar \omega}{2} \left( |S_{21}|^{2} - 1 \right),
\end{align}
where Equation (\ref{E5}) has been used. The added noise $s^{n} (\omega)$ is zero for unity power gain: a lossless, passive device. Usually, the noise power at the output is referred to the input to define a noise temperature $T_{\rm n} =  s^{n} (\omega) / |S_{21}|^{2} k$, giving
\begin{align}
\label{E9}
T_{\rm n} & \ge \frac{\hbar \omega}{2k} \left( 1 -  \frac{1}{|S_{21}|^{2}} \right).
\end{align}
For a high-gain amplifier there is a minimum noise temperature of $T_{\rm q} = \hbar \omega / 2k$, which is called the {\em standard quantum limit} (SQL). No phase preserving amplifier can have a noise temperature of less than the quantum limit \cite{Cav1}. This noise power adds to any intrinsic power from the source. If the source is also in its vacuum state, an additional half a photon of noise is added. One interpretation of the SQL is to say that at least one other source must be connected to an amplifier to provide the energy needed for amplification, and at the very least this must have vacuum fluctuations. Often, several internal sources are present, and so the SQL is not realised. To achieve the quantum limit it is prudent to choose a configuration that has the smallest number of connected degrees of freedom. A single-mode resonator parametric amplifier is a good example of this principle.

$T_{\rm n}$ is a {\em noise temperature}, but actually describes an average spectral power, $k T_{\rm n}$, and so how does it relate to noise? The main use of an amplifier is to amplify a coherent voltage or current waveform, {\em preserving phase}, meaning that the gain is the same for each of the quadrature components (time shift invariance). Equation (\ref{E3}) shows that $\hat{v}^{+}(\omega)$ is not Hermitian, and so not measurable; but combining the positive and negative frequency parts, the quadrature components of $\hat{v}(t)$ at the output, namely $ \sqrt{\hbar \omega Z_{0} / 2}   \left( \hat{b}(\omega) + \hat{b}^{\dagger}(\omega) \right) \cos (\omega t)$ and $ -i \sqrt{\hbar \omega Z_{0} / 2}   \left( \hat{b}(\omega) - \hat{b}^{\dagger}(\omega) \right) \sin (\omega t)$, are individually Hermitian, and so measurable. To calculate the fluctuation in the travelling noise-wave voltage at the output, in the absence of a signal at the input, $\Delta v_{\rm opt}(t) = \sqrt{\langle \hat{v}_{\rm opt}^{2} (t) \rangle - \langle \hat{v}_{\rm opt} (t)\rangle^{2}}$ is required. For a statistically stationary state, such as a thermal state, it can be shown through straightforward algebra that $\left( \Delta v_{\rm opt}(t) \right) ^{2} / Z_{0} = s^{n} (\omega) = k T_{\rm n} |S_{21}|^{2}$, and finally $\Delta v_{\rm in}(t)  = \sqrt{ k T_{\rm n} Z_{0} } $. The crucial point is that noise temperature is a measure of the variance of the amplitude of the noise voltage wave, and so is the relevant measure of sensitivity when an amplifier is used to amplify travelling voltage/current waveforms. This contrasts with a measurement of average power, where the fluctuation in power is the relevant measure of sensitivity. If a power detector follows an amplifier, the radiometer equation must be used: $\Delta T = T_{\rm n} / \sqrt{B \tau}$. In fact, if a signal is amplified, digitally sampled, and then analysed using an autocorrelation algorithm to give a power spectrum, the radiometer equation still holds for each spectral bin. In summary, the sensitivity of amplifiers is characterised by {\em noise temperature}, which is a spectral power, and the sensitivity of power detectors by {\em noise equivalent power}, which is a fluctuation in power for a given post-detection integration time. These are based on second-order and fourth-order statistics respectively.

\subsection{Microscopic physics}

\label{sec_mic_qua}

\begin{figure}
\noindent \centering
\includegraphics[trim = 6cm 19cm 7cm 4cm, width=50mm]{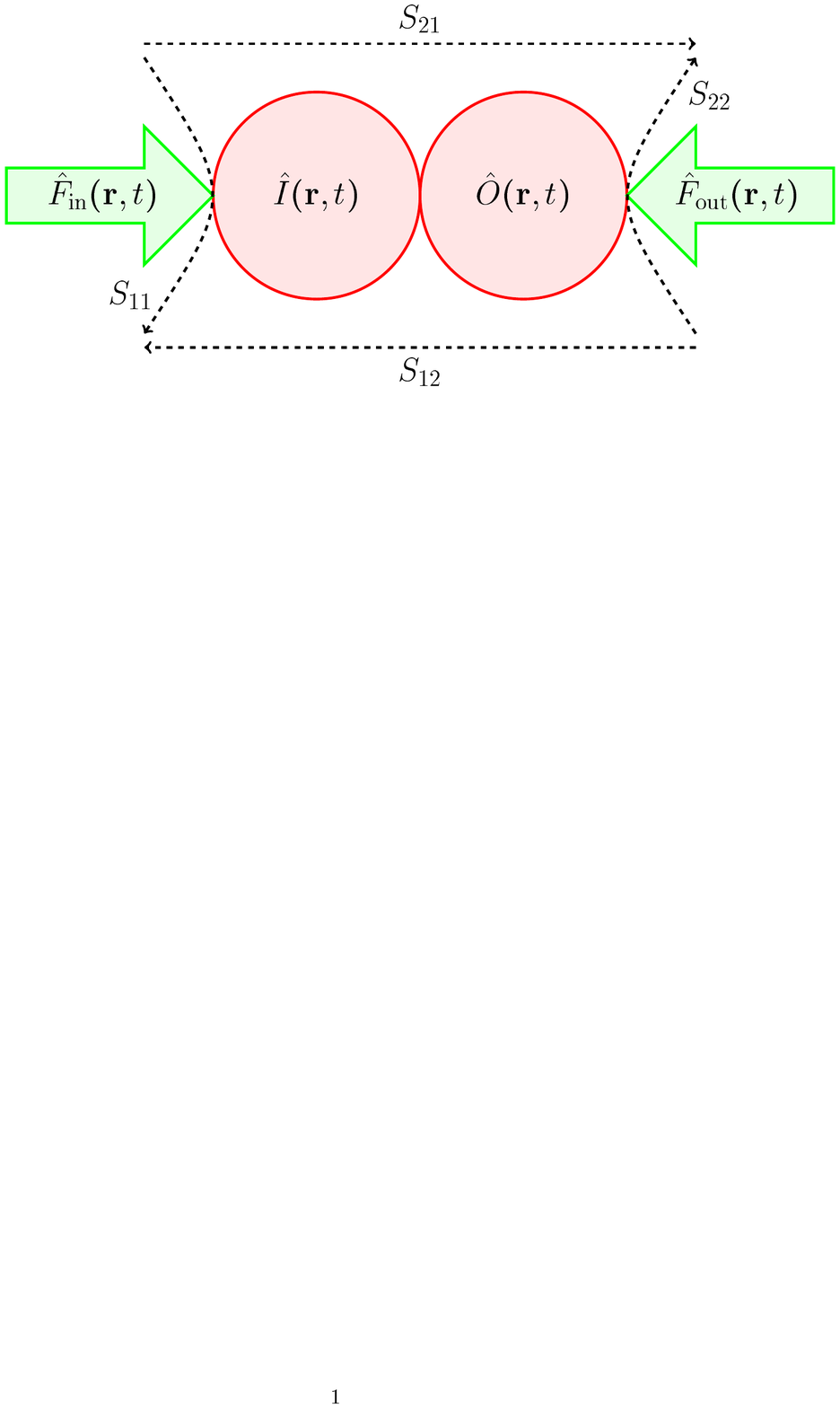}
\caption{Two generalised forces $\hat{F}_{\rm in} ({\bf r},t)$ and $\hat{F}_{\rm out} ({\bf r},t)$ act on the physical properties $\hat{I} ({\bf r},t)$ and $\hat{O} ({\bf r},t)$, respectively, of a device to create a two-port network. In the frequency domain, the quantum response functions, retarded Greens functions, are essentially the two-port scattering parameters $S_{ij}$.}
\label{figure6}
\end{figure}

It is usually sufficient to adopt a microwave systems approach to modelling, but to achieve the best possible performance, it is necessary to understand the relationship between quantum systems theory and the solid-state physics of the device. Rather than regarding the scattering parameters as coefficients, or complex probability amplitudes, it is possible to regard them as response functions in the spirit of Kubo's linear response theory. Consider the two-port network shown in Figure \ref{figure6}. There is some input quantity, $\hat{F}_{\rm in}$, such as the magnetic vector potential of an TEM wave, that couples to some property of the device $\hat{I}$, such as current density. They couple in the sense of combining to add an interaction term to the overall Hamiltonian, as in Equation (\ref{C1}). Likewise there is some output quantity $\hat{F}_{\rm out}$ that couples to some other property of the device $\hat{O}$. Various Kubo-like response formula then follow, as in Equation (\ref{C5}),
\begin{align}
\label{E1b}
\langle \Delta \hat{I}^{H}  ({\bf r},t) \rangle_{t_{0}} & = \frac{-i}{\hbar} \int_{-\infty}^{+\infty}  dt' \, \theta(t-t') \int_{\cal V} {\rm d}^{3} {\bf r}' \,  \langle \left[ \hat{I}^{I}  ({\bf r},t), \hat{I}^{\rm I} ({\bf r}',t') \right] \rangle_{t_{0}} \cdot \langle \hat{F}_{\rm in} ({\bf r}',t') \rangle_{t_{0}} \\ \nonumber
\langle \Delta \hat{O}^{H}  ({\bf r},t) \rangle_{t_{0}} & = \frac{-i}{\hbar} \int_{-\infty}^{+\infty}  dt' \, \theta(t-t') \int_{\cal V} {\rm d}^{3} {\bf r}' \,  \langle \left[ \hat{O}^{I}  ({\bf r},t), \hat{I}^{\rm I} ({\bf r}',t') \right] \rangle_{t_{0}} \cdot \langle \hat{F}_{\rm in} ({\bf r}',t') \rangle_{t_{0}}.
\end{align}
Additional steps are needed, depending on the device, to turn these expressions into scattering parameters, such as $S_{11}$ and $S_{12}$. For example, the volume integrals must be turned into surface integrals over the ports (although it may be possible to express the interaction Hamiltonian directly in terms of say the voltage at the terminals of the device, avoiding the explicit need for a volumetric formulation), and $\hat{F}_{\rm in}$ and $\hat{F}_{\rm out}$ must be described in terms of creation and annihilation operators to give travelling wave amplitudes. The central point, however, is that scattering parameters can now be identified as, essentially, Kubo response functions.

Following procedures similar to those outlined in Section \ref{sec_pow_qua}, it is possible to calculate power gain, reactive and resistive input impedance, noise generation, etc., in terms of Kubo response functions. These calculations can be carried out using the scattering parameters, but now there is a direct connection with operators that describe the quantum behaviour of the device. Remember that Kubo response functions are retarded Greens functions, which describe how a solid-state system responds to the injection of an excitation. For example, an electron or hole may by created at one space-time point and one wishes to know the complex probability of an electron or hole appearing at another point. This deep connection between quantum systems theory and device physics is compelling and highly valuable.

One important consideration relates to the distinction between backaction and noise. Noise is in a sense straightforward because it relates to the fluctuations present in outgoing waves when no external excitations are present. Backaction is more subtle because it relates to a change in the state of the applied field as a consequence of a measurement taking place. For example, the position of a particle can be measured precisely, but then all information about the momentum is lost. After this measurement, it is known where the particle is, but it is not known which way it is going. A subsequent measurement then suffers from the extreme nature of the first measurement. Often it is better not to measure the quantity of interest too precisely, so that further information can be gained at the second measurement. Electrical sensors act in the same way, and it is better for the first measurement not to constrain the the subsequent behaviour of the system too precisely. It seems that the operator $\hat{I}$ describes the way in which the amplifier feeds noise out of the input terminals, and determines the degree of backaction imposed. The
manifestation of radiated noise and backaction depends on the basis used to represent the amplifier. The travelling wave representation, which is defined only to within an arbitrary real reference impedance $Z_{0}$, presents the effects in a different way to a discrete representation where a voltage and current source are placed at the input, as is commonly done in the case of operational amplifiers \cite{Cle1}.

\subsection{Comparing performance}

\label{sec_nse_com}

Numerous fundamental physics experiments are based on measuring power spectra, but should these be carried out using low-noise detectors or low-noise amplifiers? From the perspective of sensitivity, it might be expected that the two are the same because power can be derived from voltage, and vice versa, but the measurement statistics are different, as has been seen.

When an amplifier-detector combination is used to measure power from a thermal source, the smallest change in temperature  that can be detected is given by the radiometer equation $\Delta T =  T / \sqrt{B_{\rm pre} \tau}$, where $B_{\rm pre}$ is the pre-detection bandwidth, and $\tau$ is the post-detection integration time. The smallest change in power that can be detected is then $\Delta P = k T_{\rm n} B_{\rm pre} / \sqrt{{B_{\rm pre}} \tau}$, where the Rayleigh Jeans limit is assumed when defining noise temperature. The smallest change in power that can be detected using a power detector having an intrinsic noise equivalent power of $NEP_{\rm i}$ is $\Delta P = NEP_{\rm i} \sqrt{B_{\rm pst}} = NEP_{\rm i} / \sqrt{2 \tau}$, where $B_{\rm pst}$ is the noise equivalent post-detection bandwidth. Comparing these two gives an equivalent noise temperature of
\begin{align}
\label{F1}
T_{\rm en} & =  \frac{NEP_{\rm i}}{k \sqrt{2 B_{\rm pre}}}.
\end{align}
$NEP_{\rm i}$ characterises internally generated noise, and does not include any background noise. The reason is that the noise temperature of the amplifier does not include any background noise either. If background noise is included in both cases, thereby comparing system NEP with system noise temperature, Equation (\ref{F1}) can be used. In the case of a single-pole filter having a Lorentzian profile, it can be shown that the relevant noise bandwidth is $B_{\rm pre} \approx  \pi \nu_{0} / 4 R$,  where $R = \nu_{0} / \Delta \nu$ is the spectral resolution, and $\Delta \nu$ is the full width half maximum (FWHM). Using Equation (\ref{F1})
\begin{align}
\label{F2}
T_{\rm en} & =  \left[ \frac{2 R}{k^{2} \pi \nu_{0} } \right]^{1/2} NEP_{\rm i}.
\end{align}

\begin{figure}
\noindent \centering
\includegraphics[trim = 0cm 0cm 0cm 0cm, width=140mm]{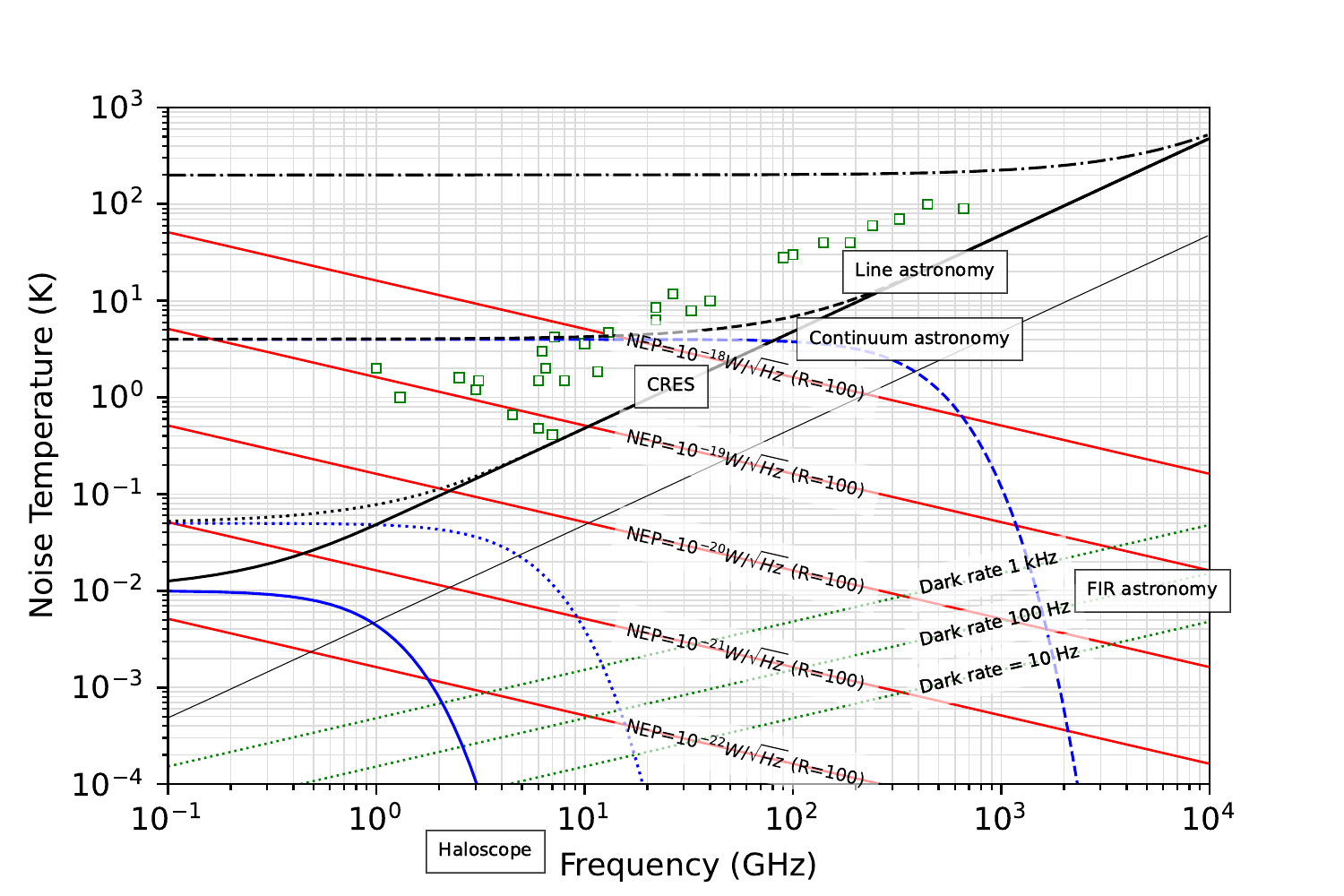}
\caption{Black lines: System noise temperature of a quantum limited amplifier preceded by a source having a physical temperatures of 10 mK (solid), 50 mK (dotted),  4K (dashed), and 200 K (dash-dot). Blue lines: Noise power radiated by single-mode sources, expressed as a noise temperature, having physical temperatures of 10 mK (solid), 50 mK (dotted) and 4 K (dashed). Red lines: Equivalent noise temperatures of detectors having noise equivalent powers of $10^{-18}$, $10^{-19}$, $10^{-20}$, $10^{-21}$, $10^{-22}$ WHz$^{-1/2}$ with $R=100$. Green dotted lines: Equivalent noise temperatures of detectors having dark photon rates of 10, 100 and 1000 Hz with $R=100$. The lowest solid faint black line shows the equivalent noise temperature of a squeezed amplifier (10 dB) operating with squeezed source at 1 mK; it can be regarded as the absolute limit of any coherent system. The squares show a range of reported noise temperatures of coherent receivers operating at a variety of temperatures. Spectral line astronomy requires coherent systems having the SQL over 100-1000 GHz. FIR infrared space based astronomy requires incoherent systems having NEPs of order $10^{-20}$ WHz$^{-1/2}$. Single electron Cyclotron Radiation Emission Spectroscopy (CRES) requires coherent systems having the SQL at 20-30 GHz. The photon production rate in Haloscopes designed for dark matter detection are extremely small, and so long integration times must be used for all realistic measurements.
}
\label{figure7}
\end{figure}

Figure \ref{figure7} shows, as black lines, system noise temperature of a quantum limited amplifier preceded by a source having a physical temperature of 10 mK (solid), 50 mK (dotted),  4K (dashed) and 200 K (dash-dot). As frequency increases, the curves converge on a single line corresponding to $\hbar \omega / k$. Ordinarily, this indicates the sensitivity limit of coherent receivers. The SQL increases with $\nu_{0}$, but does not depend on bandwidth. A 10 mK source (typical of a dilution refrigerator) allows the SQL to be achieved at frequencies down to about 500 MHz. The dashed curve corresponds to a 4K source (typical of a pulse tube cooler), and shows that the SQL can be achieved down to 100 GHz. SIS mixers (see later) approach the SQL over the range 100 GHz to 1 THz. An upward looking radiometer or a space based radiometer always has the ~3 K CMBR as its background, and so there is no benefit in using quantum-noise-limited amplifiers below 10 GHz. The dash-dot line, corresponding to a 200 K source, is essentially the temperature seen by an Earth Observation instrument; it is clear that uncooled amplifiers are suitable for most remote sensing applications. The equivalent noise temperatures of detectors having noise equivalent powers of $10^{-18}$, $10^{-19}$, $10^{-20}$, $10^{-21}$, $10^{-22}$ WHz$^{-1/2}$ with $R=100$ are shown in red. The equivalent noise temperature improves as the centre frequency $\nu_{0}$ is increased, and as the resolution $R$ is lowered. The blue lines show the noise power radiated by single-mode sources, expressed as noise temperature, having physical temperatures of 10 mK (solid), 50 mK (dotted) and 4 K (dashed).  A 10 mK source allows detectors having NEPs of better than $5 \times 10^{-21}$ WHz$^{-1/2}$ to be exploited down to 1 GHz: for R=100. There is no point in using a detector of any kind having a noise temperature that is significantly below the background limit.  The plot shows that ultra-low-noise FIR power detectors, NEP$\approx 10^{-20}$ WHz$^{-1/2}$, can be used over 1-10 THz, where the blackbody spectrum of the CMBR falls away steeply. In cases such as these, where the noise is dominated by intrinsic detector noise, it can be beneficial to have a number of optical modes available for absorbing signal power, as described in Section \ref{sec_mul_det},
motivating the use of multimode detectors in space-based FIR astronomy.

Figure \ref{figure7} shows why power detectors, often called {\em incoherent receivers}, are used for high frequency low-resolution measurements, whereas amplifiers, often called {\em coherent receivers}, are used for low frequency high-resolution measurements. The crossover occurs at millimetre wavelengths, where the two approaches have similar sensitivities. The squares show a range of state-of-the-art noise temperatures of coherent receivers operating at a variety of physical temperatures. The crucial point is that the various technologies all fall short of the SQL by a factor of a few, but track it with frequency. The plot suggests that coherent receiver technology starts to suffer from radiometric leakage at frequencies below 10 GHz even though the physical temperature is often lower. Controlling inadvertent stray light in a cryostat can be surprisingly challenging. The best detectors have NEPs of $10^{-19}$ to $10^{-20}$ WHz$^{-1/2}$ showing that two orders of magnitude improvement at microwave frequencies would be highly beneficial, leading to instruments that are far superior to even squeezed amplifiers for certain low-spectral-resolution (R=100) applications. Incoherent receivers can also suffer badly from radiometric leakage.

It should not be assumed that the SQL, $T_{\rm q}$, cannot be beaten. This limit exists only in the case of phase preserving amplifiers. Some amplifiers, however, get the energy needed for amplification from a coherent pump tone, and can be engineered to have different gains for the in-phase and out-of-phase components. The noise temperature of one quadrature can be lowered below the SQL, but only at the expense of the noise temperature of the other: as required by Heisenberg's uncertainty principle. This process is called {\em squeezing}, and squeezing factors, gain ratios, of 10-15 dB are achievable. Higher squeezing factors are challenging because any phase imperfections, which may be time dependent, degrade the fidelity of the effect. The circle in Figure \ref{figure5} becomes an ellipse, and the higher the squeezing factor, the higher the sensitivity to changes in the orientation of the ellipse. The most sophisticated systems squeeze both the source, and the noise from the the amplifier, allowing exceptionally sensitive measurements to be made.  The faint solid line is the SQL of a 10 dB squeezed system at 1 mK, showing that lower noise temperatures are possible if this exotic mode of operation can be achieved and developed for practical applications

\section{Superconducting devices and circuits}

Superconducting thin-film devices provide an excellent technological platform for exploiting concepts in quantum sensing. When a BCS superconductor is cooled below its critical temperature $T_{\rm c}$, an energy gap forms, $E_{\rm g} = 7 k T_{\rm c} /2$, and the material's  electrical, magnetic and thermal characteristics change significantly. Most quantum devices operate at $T_{\rm p} \approx 0.1 \, T_{\rm c}$, but a few operate at higher temperatures, $T_{\rm p} \approx T_{\rm c}$. Some of the most important materials, deposited using ultra-high-vacuum sputtering, and patterned using ultraviolet lithography, are listed in Table \ref{Super}. The gap frequency $ f_{\rm g} = E_{\rm g} / h$ is important because below $f_{g}$ the material has near-zero electrical resistivity, whereas above $f_{g}$ superconducting pairs are broken to create single-electron excitations, quasiparticles, which have appreciable resistivity. Power detectors and photon counters usually operate at frequencies above $f_{\rm g}$, and extend up through the infrared and X-ray regions. Amplifiers, frequency convertors and transmission lines must avoid breaking pairs and so operate from kHz to fractional THz frequencies. Reactively sputtered materials such as NbN and NbTiN are popular because they allow operation to around 1.2 THz. It is also routine to fabricate devices using multilayers (TiAu, MoAu, TiAl). Although the films do not diffuse and stay physically distinct, a proximity effect causes superconducting quasiparticles and pairs to leak, and the multilayer behaves as a homogeneous superconductor having properties that are intermediate between those of the constituent layers. For example, $T_{\rm c}$ can be adjusted over the range 50-500 mK to a precision of about 5 mK. A long range lateral proximity effect also occurs, where a wiring contact can, without material diffusion, change the properties of the active device. Table \ref{Devices} lists a number of important device types, and indicates whether they rely on pair breaking or pair preservation.

\begin{table}
\begin{center}
{\begin{tabular}{ccccc}
  Material & $T_{\rm c}$ (K) & $E_{\rm g}$ (meV) & $f_{g}$ (GHz) \\  \hline
  NbN & 16 & 4.8 & 1160  \\
  Nb & 9.3 & 2.8 & 680 \\
  Ta & 4.48 & 1.35 &  325 \\
  Al & 1.2 & 0.36 & 90 \\
  Mo & 0.9 &0.27 & 65 \\
  Ti & 0.39 & 0.11 & 26 \\
  \hline
\end{tabular}}
\end{center}
\caption{Illustrative characteristics of superconductors commonly used to fabricate quantum sensors. $T_{\rm c}$ is the critical temperature, $E_{\rm g}$ the energy gap, and $f_{g}$ the associated pair breaking frequency.}
\label{Super}
\end{table}

\begin{table}
\begin{center}
{\begin{tabular}{ccc}
  Material & Pair breaking  &  Wavelength range \\  \hline
Passive & No & Microwave to submm \\
    SQUID & No & RF   \\
    SIS & No &  Submm  \\
    TES & Yes &  Submm, FIR, Optical and X-ray   \\
    KID & Yes &  Submm, FIR   \\
    Paramp & No & Microwave, MMwave   \\
    SPNWD & Yes & Optical
\end{tabular}}
\end{center}
\caption{Various superconducting devices used in fundamental physics experiments. Descriptions are given in the text. The mode of operation and typical operating wavelength are listed. }
\label{Devices}
\end{table}

In this short overview, it is not possible to list all of the superconducting components available, but it is useful the illustrate the breadth of the technology available. It should also be appreciated that many of the device types described below can be combined to create complex microcircuits having a high degrees of functionality: large format imaging arrays and chip spectrometers.

\begin{figure}
\noindent \centering
\includegraphics[trim = 1cm 1cm 1cm 1cm, width=70mm]{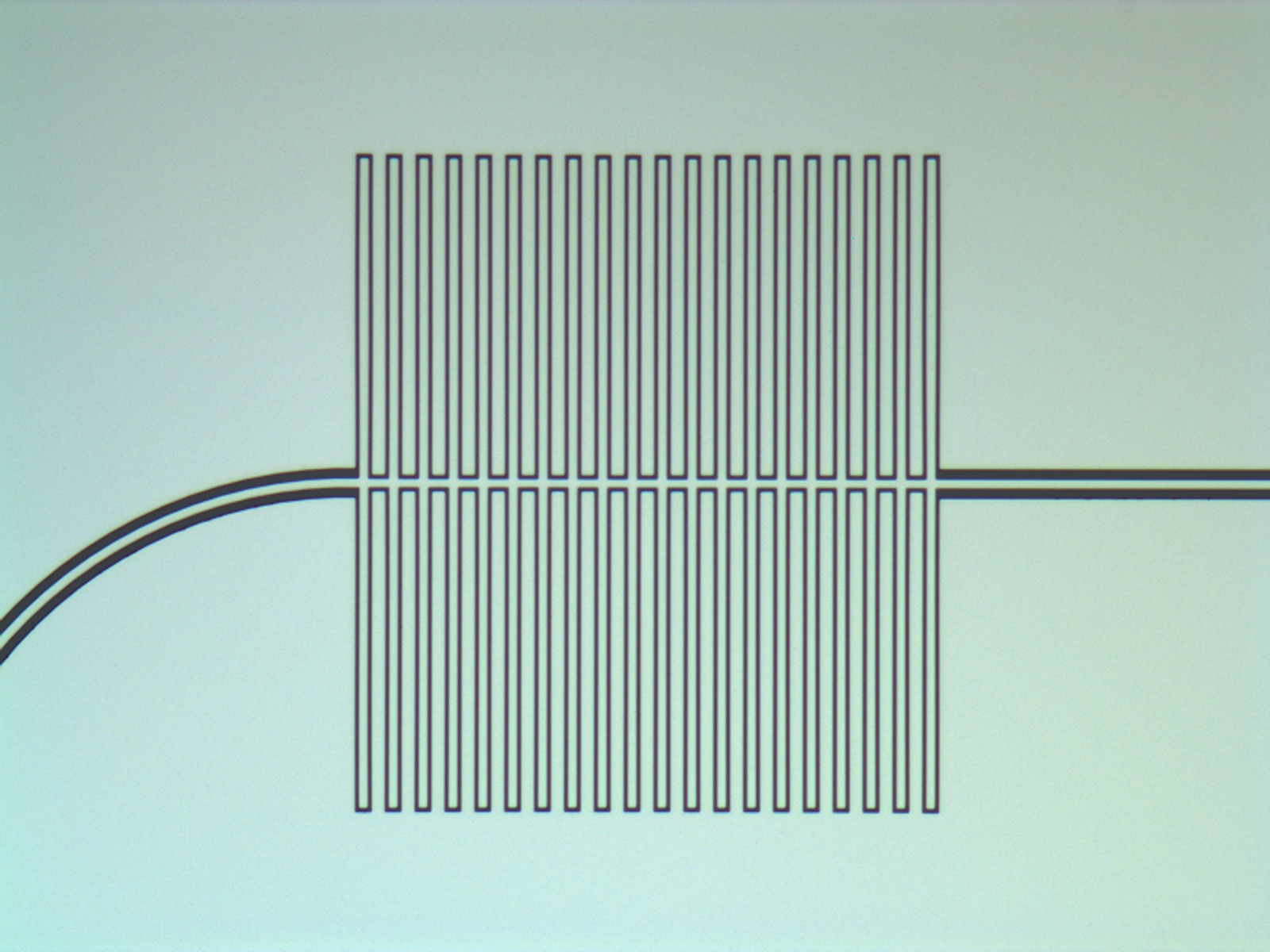}
\caption{Millimetre-wave thin film superconducting filter based on coplanar transmission line. Fabricated by the Quantum Sensors Group in Cambridge (see Acknowledgements).}
\label{figure8}
\vspace{2mm}
\includegraphics[trim = 1cm 1cm 1cm 1cm, width=70mm]{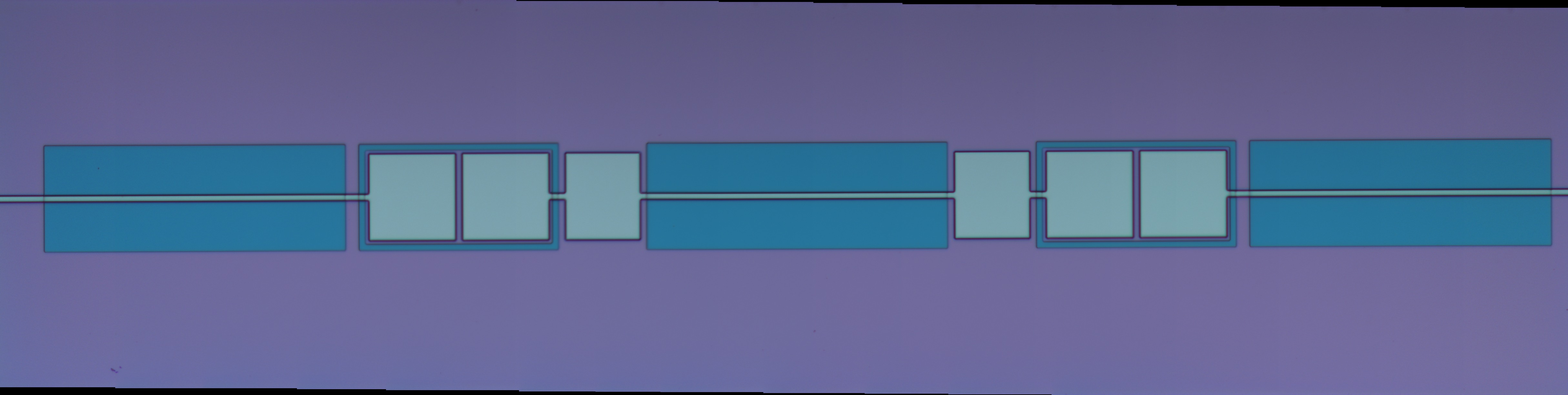}
\caption{Millimetre-wave superconducting filter based on parallel capacitors and series capacitors and inductors. The films are typically 100 nm thick. Fabricated by the Quantum Sensors Group in Cambridge (see Acknowledgements).}
\label{figure9}
\end{figure}

{\bf Passive components:} RF components, such as micron-scale transmission lines, directional couplers, filters and loads can be realised in microstrip ($ 5 < Z_{0} <40 \, \Omega$)  and coplanar ($ 70 < Z_{0} <150 \, \Omega$) configurations. These structures use 50-500 nm thick superconducting, normal metal and dielectric films, such as SiO and SiO$_{2}$, to create passive components that can operate to 1.2 THz: Figure \ref{figure8} and Figure \ref{figure9}. Superconducting RF components have a number of advantages: (i) For frequencies below $f_{\rm g}$, and at low powers, the films are essentially lossless enabling exceptional behaviour. Dielectric loss in deposited and surface oxides then becomes the biggest dissipative factor, and amorphous dielectrics lead to troublesome Two Level System (TLS) noise. In fact, the deposition and control of low-loss oxides is one of the biggest challenges facing the technology.(ii) The surface impedance of a superconductor is complex valued. The real part is caused by dissipation, usually in the form of quasiparticle scattering, and the imaginary part by reactive energy stored in the inertial behaviour of undamped pairs. If the cross section of the dielectric region is sufficiently small, the energy stored in the kinetic inductance of the film can be comparable with the energy stored in the electromagnetic field. The consequential reduction in wavelength results in devices being physically smaller than would otherwise be the case. (iii) Superconducting thin-film transmission lines are only lossless for powers below, roughly speaking, -50 dBm. At higher powers the quasiparticle population increases and heats due to the photon absorption rate being greater than the quasiparticle recombination rate, and the surface impedance changes. Superconducting resonators show a rich variety of behaviour because on tuning through the resonance the stored energy changes, modifying the equivalent circuit parameters: the resonant frequency and quality factor of the underlying resonance depend on the frequency and strength of exciting tone \cite{Cnt1}.

{\bf Superconducting Quantum Interference Device:} SQUIDs were the first superconducting sensors to be used for science. They operate well below $T_{\rm c}$ because of the need to maintain a long-range coherent superconducting state \cite{Sqd1}. Imagine a closed loop of superconducting material. The line integral of the magnetic vector potential is equal to the flux enclosed, but according to the Aharanov-Bohm effect, the line integral contributes a phase factor to the bosonic wavefunction of the superconducting pairs. Because the phase around a closed loop must be single valued, only certain values of flux are allowed to exist inside the loop. The quantum of magnetic flux is $ \Phi_{0} = h / 2 e = 2.1 \times 10^{-15}$ Tm$^{-2}$. A DC SQUID comprises a superconducting ring in which two tunnel junctions are inserted on opposite sides. If the ring is current biased through two additional connections, arranged to give a symmetric configuration, the voltage across the tunnel junctions provides a measure of the screening current in the ring. The voltage is then periodic as an external signal flux is applied and individual quanta enter the ring. This device can be used for extremely sensitive field measurements (fT Hz$^{-1/2}$). Another level of sophistication uses the amplified voltage to feed back flux into the ring through a thin-film transformer. The feedback holds the total flux constant, and the feedback voltage gives a linear measure of the externally applied flux. Finally, a low-inductance input transformer can be added to create an ultra-low-noise current to voltage convertor. Noise currents of pA $\sqrt{\rm Hz}$ are routinely achieved. SQUIDS have been developed for applications such as biological and biomedical magnetometry, geology, and even oil exploration. SQUIDs are also be used for reading out and multiplexing Transition Edge Sensors.

{\bf Superconducting Parametric Amplifier:} SPAs can be based on the nonlinear behaviour of SQUIDs by modulating the flux in the ring with an external RF pump source, or on the nonlinear behaviour of thin-film transmission lines \cite{Prm1}. In both cases, they must be used below $f_{\rm g}$. In the case of transmission lines, the signal is combined with a high-level pump tone, which modulates the parameters of the device and transfers energy to the signal, resulting in gain (10-20 dB). Half-wavelength superconducting resonators make excellent narrow band amplifiers at microwave frequencies, and are predicted to work well at millimetre wavelengths \cite{Prm2}. The advantages of resonators are that only small pump powers are needed, keeping phase noise low, and the number of degrees of freedom can be controlled, minimising the number of modes that can contribute to vacuum fluctuation noise. The bandwidths of resonators are small $R \ge 200$, and so for broadband applications $R \le 5$ travelling wave structures are needed. The difficulty with travelling wave devices is that the cross section of the transmission line must be small (50 nm thick films, 1-2 $\mu$m wide) to maximise the kinetic inductance fraction, but the lengths must be large (0.5 m) to maximise gain. It is difficult to achieve uniform, defect free fabrication, and there are other complications associated with dispersion engineering and harmonic suppression. Also, relatively large pump powers are needed, leading to heating. Superconducting films display both resistive and inductive nonlinearities, and these contribute simultaneously to the operation of amplifiers based on transmission lines \cite{Prm2}. Additionally, at least two non-linear mechanisms are present (gap moduation and quasiparticle generation), and these have different speeds and power thresholds. Overall, SPA's achieve exceptional behaviour at microwave frequencies, frequently approaching the quantum limit. Most configurations give phase-preserving amplification, but the intrinsic ultra-low-noise behaviour has also allowed squeezing to be demonstrated.

{\bf Superconductor Insulator Superconductor mixer:} SIS mixers were the first superconducting devices to find widespread use in astrophysics \cite{Sis1,Sis2}, and now form a technological cornerstone of high resolution ($R \approx 10^{8}$) submillimetre-wave spectral line astronomy. As discussed in Section \ref{sec_nse_com}, high-resolution instruments favour coherent systems. SIS mixers exploit the complicated dynamics of quasiparticle tunnelling in dielectric barriers, but it is necessary to suppress Josephson pair tunnelling by the application of a small DC magnetic field in the plane of the barrier. A typical tunnel junction has an area of 1 $\mu $m$^{2}$, to minimise capacitance, allowing near quantum-noise-limited down conversion from submillimetre (100 GHz to 1 THz) to microwave (1-10 GHz) frequencies. SIS mixers are fascinating devices because they bridge the gap between classical mixers, based on notions of IV curve nonlinearity, and photon-energy convertors, based on notions of creation and annihilation operators acting on field states. As the frequency of the LO is increased, a changeover in behaviour occurs when the photon energy becomes greater than the scale size of the nonlinearity in the IV curve ($\approx$ 100 GHz). Photon-assisted tunneling steps appear due to the quasiparticle states on one side of the barrier being energy(essentially frequency) modulated. The change from classical to quantum behaviour brings gain on down conversion (in contrast to classical mixers which must have a 3 dB loss at best), quantum-limited noise temperature (due to the presence of the LO), and the appearance of quantum capacitance due to mismatched quasiparticle states on either side of the barrier creating a sloshing probability current. SIS mixer technology has enabled pioneering submillimetre-wave telescopes to be built, such as the James Clerk Maxwell Telescope in Hawaii, and the Atacama Large Millimeter Array in Chile, and has been flown to Lagrange Point 2 on the Herschel Space Telescope.

\begin{figure}
\noindent \centering
\includegraphics[trim = 0cm 0cm 0cm 0cm, width=60mm]{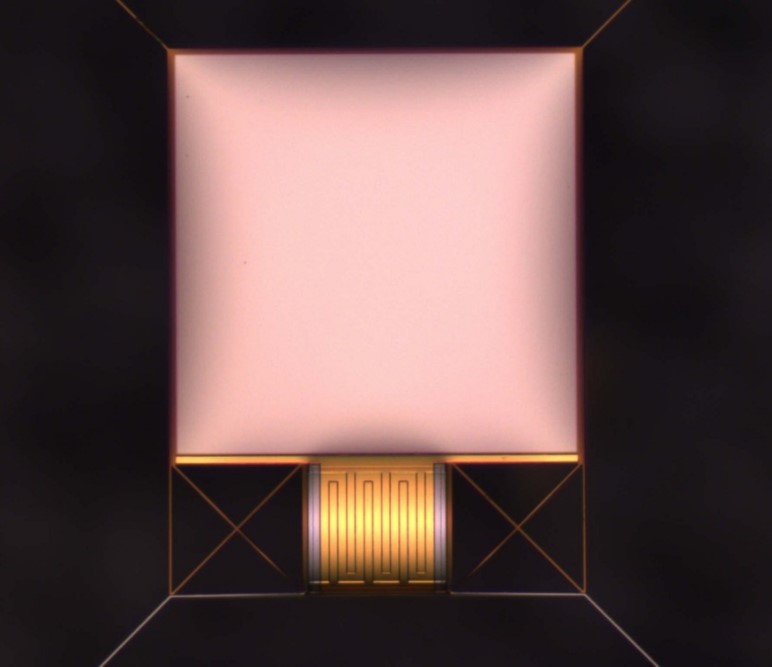}
\caption{Free-space far-infrared superconducting transition edge sensor. The superconducting MoAu bilayer at the bottom of the picture has gold bars to suppress noise. It is fabricated on a 200 nm thick SiN membrane, which is supported by legs which are 200 nm thick, 1 $\mu$m wide and can be up to 1 mm long. Nb wiring runs out along the two legs at the bottom of the picture. The infrared absorber comprises a few nm of disordered $\beta$-phase Ta. Fabricated by the Quantum Sensors Group in Cambridge (see Acknowledgements).
}
\label{figure10}
\end{figure}

\begin{figure}
\noindent \centering
\includegraphics[trim = 0cm 0cm 0cm 0cm, width=70mm]{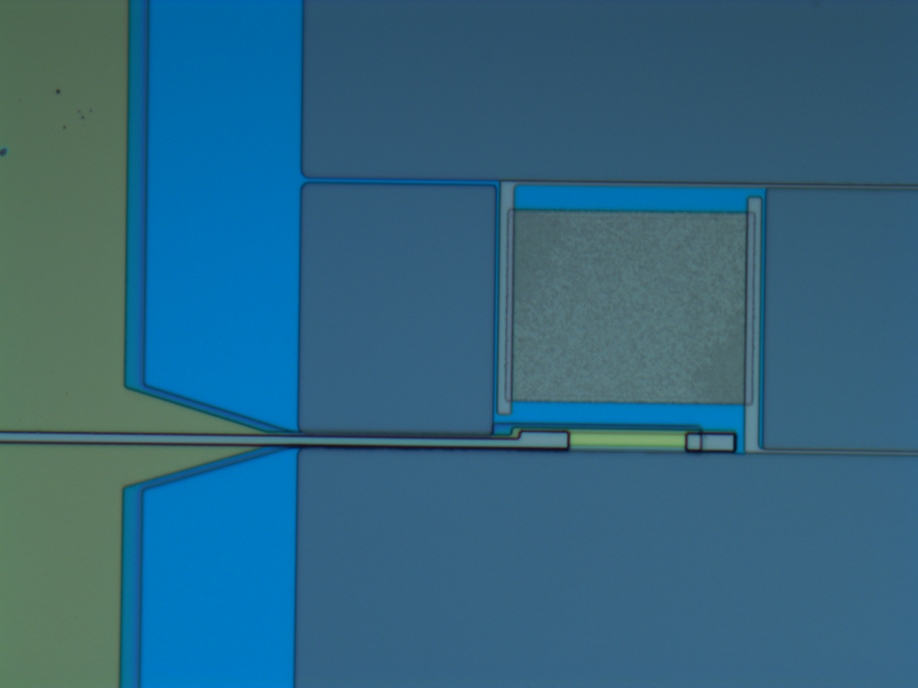}
\caption{Microstrip coupled millimetre-wave transition edge sensor. The primary devices is a TiAl bilayer supported on a 200 nm SiN membrane. The legs are 4 $\mu$m wide and support Nb wiring and a superconducting microstrip transmission line, which is fabricated in Nb with SiO$_{2}$ insulator, is terminated in a 20 $\Omega$ gold/copper load. Fabricated by the Quantum Sensors Group in Cambridge (see Acknowledgements).
}
\label{figure11}
\end{figure}

{\bf Transition Edge Sensor:} TESs have been the work horse of submillimetre-wave astronomy for many years, particularly for mapping spatial variations in the intensity and polarisation of the CMBR, revealing multipole acoustic oscillations in the plasma of the early Universe. As discussed in Section \ref{sec_nse_com}, low-resolution instruments favour incoherent systems. The basic device comprises a superconducting film isolated from the heat bath by either SiN legs (200nm thick and 2 $\mu$m wide) or by judicious engineering of electron-phonon decoupling in the superconductor. When the superconducting film is connected to a low impedance (m$\Omega$) voltage source, electrothermal feedback causes the film to self bias on its superconducting transition. External radiation is then applied optically to a nearby absorbing film made of a different superconductor, Figure \ref{figure10}, or to a load that terminates a superconducting microstrip transmission line, Figure \ref{figure11}. When energy is absorbed, electrothermal feedback holds the operating point constant by swapping optical power for bias power. The bias current falls and is read out using a SQUID. Electrothermal feedback causes a TES to respond more quickly than the open-loop thermal time constant would suggest. TESs have been developed extensively for most of the electromagnetic spectrum, and although various noise mechanisms are present they can be suppressed to the point where the phonon shot noise in the thermal isolation dominates, giving NEPs of 10$^{-17}$ to 10$^{-20}$ WHz$^{-1/2}$. TESs can be assembled into very large arrays, and the microstrip versions have been combined with superconducting RF components to make chip spectrometers for astronomy and Earth Observation. TESs are also been used at optical wavelengths for laser interferometry, dark matter searches, and have been developed into a sophisticated technology for far-infrared and X-ray space telescopes.

{\bf Kinetic Inductance Detector:} KIDs are being developed to replace TESs in applications where very large format imaging arrays are needed \cite{Kid1}. They can been used from submillimetre to X-ray wavelengths, with time-resolved photon-counting spectroscopy being possible at the shortest wavelengths. A low-power coherent tone is applied to a superconducting microwave resonator so that its complex transmission factor can be monitored. The resonator can be a distributed transmission line, or a discrete circuit that takes the form of an optical pixel. When signal power, or indeed an individual signal photon, is absorbed by the material of the resonator, the surface impedance and resonant frequency change, and this modulates the microwave transmission amplitude and phase. The real promise of this device is that thousands of pixels can be weakly coupled to a single superconducting readout line, and a densely packed comb of microwave tones generated digitally. The output signal can then be sampled, and a real-time FFT used to measure the complex transmission factors of all of the devices simultaneously.  The ambition is to create submillimetre-wave and far-infrared cameras having tens of thousands of pixels. Various trade offs have to be considered, but NEPs ranging from 10$^{-17}$ WHz$^{-1/2}$ to 10$^{-20}$ WHz$^{-1/2}$ have been achieved. A challenge with these devices is to ensure that optical behaviour can be maintained, in terms of beam patterns and efficiency, whilst not degrading the microwave response, such as responsivity and noise. These devices are subject to the complicated dynamics of superconducting resonators, and the generation of quasiparticles by the readout tone is a particular consideration. It is difficult to optimise the optical and readout characteristics simultaneously at frequencies much below 100 GHz, because the signal needs to break pairs but the readout needs to preserve pairs.

The above list of devices is certainly not exhaustive. TESs operate at $T_{\rm c}$, which is usually chosen to be twice the bath temperature $T_{\rm b}$, and although $T_{\rm b}$ is typically in the range 50-300 mK, the active part of the device is not as cold as it might be, leading to noise. The Cold Electron Bolometer (CEB) is an ingenious device that overcomes this problem \cite{Ceb1}, enabling NEP's of 10$^{-21}$ WHz$^{-1/2}$ to be achieved. Also, Superconductor Nanowire Single Photon Detectors (SNSPD) are being used for optical photon counting \cite{Had1}, and Superconducting Qubits for microwave photon detection \cite{Qub2}. The squares in Figure \ref{figure7} show the noise temperatures of a range of different coherent receiver technologies. Again, this is far from exhaustive, but it can be seen that there is a particular need to further develop amplifiers, and solid-state squeezed systems for frequencies in the range 1-100 GHz. Additionally, there is a need to develop ultra-low-noise incoherent receivers for the whole of the microwave-FIR region: 0.1 GHz to 10 THz.

\section{Concluding Remarks}

A new generation of ultra-low-noise sensors is required. The systems and their associated devices must push at quantum limits and so must be designed using quantum mechanical methods. There is a particular need for detectors, amplifiers, frequency convertors and imaging arrays for radio to infrared wavelengths, where existing devices fall short of theoretical limits. In some cases, the needed advances will be achieved through refinements in existing technology, but in other cases new device types must be invented. Crucially, raw sensitivity is rarely sufficient, and other characteristics such as quantum efficiency, bandwidth, linearity, saturation power and stability must obtained simultaneously. One of the biggest challenges is to achieve artefact-free behaviour at the quantum level, particularly when an instrument is to be used in a harsh environment or flown in space. The needed innovations go beyond engineering methods and relate to the the development of theoretical and numerical tools. Quantum information theory, quantum field theory, device physics and classical circuit theory must be brought together, and described using a common language, if the quantum sensors challenge is to be tackled in a robust way.

\section*{Acknowledgements}

I am grateful to UKRI/STFC for the awards Quantum Technology for Measurement of Neutrino Mass (QTNM) ST/T006307/1, Quantum Sensors for the Hidden Sector (QSHS) ST/T006625/1, and Ultra-low-noise Superconducting Spectrometer Technology for Astrophysics ST/V000837/1. Over the years I have had numerous enlightening discussions with colleagues on superconducting device physics. In particular, I would like to thank Christopher Thomas, David Goldie, Songyuan Zhao, Michael Crane and Dorota Glowacka for their exceptional work on developing, fabricating and testing the many devices studied by the Quantum Sensors Group in Cambridge over  a period of 20 years. I would also like to thank Dennis Molloy and David Sawford for their outstanding work on engineering and operating a long list of ultra-low-noise cryogenic systems.

\section*{Biography}

Stafford Withington has worked on ultra-low-noise experiments for astronomy and fundamental physics for many years, including the development of ultra-low-noise instruments for submillimetre-wave and far-infrared space-based applications. His quantum sensors group at Cambridge has been developing and fabricating superconducting devices, microcircuits and imaging arrays for over 20 years. Stafford is now Emeritus Professor of Physics at the University of Cambridge, and Visiting Professor and Senior Researcher in the Department of Physics at the University of Oxford. He has held fellowships at Downing College Cambridge, All Souls College Oxford, Queens College Oxford and a Royal Society Fellowship at Chalmers University Sweden. He worked for various companies early in his career, including Ferranti Electronics Ltd., Marconi Space and Defence Systems Ltd., and Rolls Royce aircraft Engines (1971) Ltd.

\end{document}